\documentclass[superscriptaddress,twocolumn]{revtex4-1}
\usepackage[latin1]{inputenc}
\usepackage{amsmath}
\usepackage{amsfonts}
\usepackage{amssymb}
\usepackage{graphicx}
\usepackage{appendix}
\usepackage{dcolumn}
\usepackage{subfigure}
\usepackage{float}
\usepackage{bm}
\usepackage{latexsym}
\usepackage{epsfig}
\usepackage[usenames]{color}
\usepackage[colorlinks=true]{hyperref}
\def\lsim{\mathrel{\rlap{\lower4pt\hbox{\hskip1pt$\sim$}}
    \raise1pt\hbox{$<$}}}                
\def\gsim{\mathrel{\rlap{\lower4pt\hbox{\hskip1pt$\sim$}}
    \raise1pt\hbox{$>$}}}                

\begin{document}

\title{Revealing Non-circular beam effect in WMAP-7 CMB maps with
BipoSH measures of Statistical Isotropy}

\author{Nidhi Joshi}
\email[e-mail: ]{nidhijoshi@ctp-jamia.res.in}
\affiliation{Centre for Theoretical Physics, Jamia Millia Islamia, New Delhi 110025, India}
\affiliation{IUCAA, Post Bag 4, Ganeshkhind, Pune-411007, India}

\author{Santanu Das}
\email[e-mail: ]{santanud@iucaa.ernet.in}
\affiliation{IUCAA, Post Bag 4, Ganeshkhind, Pune-411007, India}

\author{Aditya Rotti}
\email[e-mail: ]{aditya@iucaa.ernet.in}
\affiliation{IUCAA, Post Bag 4, Ganeshkhind, Pune-411007, India}

\author{Sanjit Mitra}
\email[e-mail: ]{sanjit@iucaa.ernet.in}
\affiliation{IUCAA, Post Bag 4, Ganeshkhind, Pune-411007, India}

\author{Tarun Souradeep}
\email[e-mail: ]{tarun@iucaa.ernet.in}
\affiliation{IUCAA, Post Bag 4, Ganeshkhind, Pune-411007, India}

\begin{abstract}

Mild, unavoidable deviations from circular-symmetry of instrumental
beams in current Cosmic Microwave Background (CMB) experiments
pose a significant challenge to deriving high precision inferences
from the high sensitivity and resolution of CMB measurements. It is
important to be able to measure and characterize this subtle effect
since it has bearing all subsequent inferences, including the
cosmological parameters. We present analytic results, reconfirmed by
numerical simulations, that CMB maps of cosmological signal that
respect underlying statistical isotropy (SI) symmetry, measured with
an instrument that has mildly non-circular (NC) beam would,
nevertheless, exhibit SI violation. We show that
appropriate observable measures constructed within the Bipolar
Spherical Harmonic (BipoSH) representation of SI violation capture
subtle NC-beam effects coupled with the scan strategy of the
instrument.

Accompanying their 7-year data release, the WMAP team published
very high significance measurements of non-zero BipoSH spectra,
$A^{20}_{\ell \ell}$ and $A^{20}_{\ell-2 \ell}$, in the ``W'' and
``V'' band of the experiment. The BipoSH measurements at the two
frequency channels show significant differences that point against an
origin in the achromatic cosmological CMB anisotropy signal. 
We present a strong case that the quadrupolar components of the NC-beams,
$b_{l2}$, of the respective channels have the most dominant contribution towards BipoSH measurement and create non-trivial qualitative features of the measured BipoSH spectrum that may be difficult to mimic through other effects.
The fact that subtle levels
of non-circularity, e.g., $|b_{l2}|/b_{l0} \lsim 0.01$ as seen in WMAP beams,
lead to measurable BipoSH spectra points to the immense promise and
potential of the BipoSH representation.

The scan strategy of CMB experiments typically employ multiple visits with varying
orientation at different sky positions. Hence, the effective smoothing
associated with the NC part of the beam is not expected to be
represented by the instantaneous NC-beam response. In general, the effective beam at
any point will also have variations over the sky. Using the BipoSH measurements it is possible to construct an equivalent single hit,`parallel transported' {\em effective}
NC-beam that matches the angular power spectrum as well as provides
a good estimate of
violation of SI in the observed maps.
Thus, using the BipoSH formalism, we successfully characterize
an important systematic effect that limits all attempts to probe the
SI assumption implied in the cosmological principle at high precision
through CMB measurements beyond the angular power spectrum.

\end{abstract}
\maketitle

\section{Introduction}

Cosmic Microwave Background (CMB) experiments with high sensitivity,
fine angular resolution covering almost the full sky have reached a
point where even mild, unavoidable, deviations from circularity of the
instrumental beam around its pointing direction poses a challenging
systematic effect that must be overcome within finite computational
resources to fully realize potential of these measurements for
precision cosmology.

The effect of non-circular (NC) beam on angular power spectrum of CMB
has been studied in literature and the non-trivial impact on high
precision cosmological inferences has been appreciated but not
satisfactorily resolved, particularly, within available computational
resources. A significant body of literature attempting to deal with NC
beam effect on the angular power spectrum exists, E.g.,
~\cite{TS-BR,PF-OD-FB,SM-AS-TS,TS-SM-AS-SR-RS,FebeCoP2011,fastconv2001,WMAP_GH2006,Planckmap_Ash2011}.
Further, the beam response function can acquire an inherent time
dependence which complicates and hinders the deconvolution
enormously. Consequently, beam imperfections, coupled with the scan
strategy, lead to very complex modification of the signal demanding
high computational resources to assess the final effect on the
estimation of angular power spectrum and the cosmological
parameters.
It is important to be able to capture the NC-beam effect in observed
CMB maps.  We provide a computationally efficient characterization of
the NC-beam effect by focusing on its effect beyond the angular power
spectrum.  Beyond the angular power spectrum, the improvement in
quality of CMB measurements now allow observational confrontation of
simplifying, theoretical, assumptions implicit in current
cosmology. Given CMB fluctuations are a realization of correlated,
Gaussian random field, the two point correlation function uniquely
describes the statistics of the fluctuations. Statistical Isotropy~(SI) implies rotational
invariance of the two point correlation function. Breakdown of SI can
be parametrized by the expansion coefficients of the two point
correlation function in Bipolar spherical harmonic (BipoSH) basis
known as BipoSH coefficients.  Cosmological CMB temperature
fluctuations are generally assumed to be a realization of statistically isotropic,
Gaussian, correlated, random field on the sphere, consequently, the
angular power spectrum has been the primary observational target of
most CMB experiments. However, current and upcoming CMB experiments
also hold the promise to observationally constrain the underlying,
often implicit, SI assumption (closely linked to the so called,
`cosmological principle') employed in cosmology.

SI assumption has
been under intense scrutiny with hints of various `anomalies'
persisting in successive years of WMAP
data~\cite{MT-OC-AH,PB-KG-AB,CC-DH-DS,KL-JM,HE-FH-AB}. Violation of SI
can be generated both, from theoretically motivated possibilities such
as, cosmic topology, anisotropic cosmologies etc.  as well as from
observational artifacts, such as, beam non-circularity, anisotropic
noise, foreground residuals,
etc.~\cite{ML-JL,NC-DS-GS,BPS2000,JL,LA-SM-MB,AG-CC-MP,TS2006,MA-TS,AP-MK,AR-MA-TS}. Searching
for the most plausible candidate of SI violation is a non-trivial
endeavor which strongly depends on observational and theoretical hints
to narrow down the possibilities. While NC-beam effects do
pose as a serious systematic contaminant for SI measurements, we
demonstrate that the BipoSH representation measures of SI violation
are extremely effective in capturing, characterizing, and
possibly, isolating subtle NC-beam effect in CMB maps.

The recent detection of a quadrupolar power asymmetry `anomaly' in
WMAP 7 year data (WMAP-7) by the WMAP team using BipoSH representation
measures ~\cite{CB-RH-GH} has proved to be an intriguing observation,
that is yet to be satisfactorily explained. The published detections
were significant both in the V-Band and W-Band frequency maps. Due to
mildly significant differences in the BipoSH signal at the two
frequencies and the fact that the signal seems to exhibit azimuthal
symmetry in the ecliptic coordinate, it was suspected not to be of
cosmological origin. Effects of residual galactic foreground emission
would, arguably, be expected to have associated symmetries in the
Galactic coordinate.

Observed CMB anisotropy (and polarization) on the sky is a
convolution of the underlying cosmological CMB signal with the
instrumental beam response function.  The instrumental beams in most
CMB experiments are nearly circularly (azimuthal) symmetric, but mild
NC deviations do arise due to unavoidable limitations in instrumental
design, function and fabrication, e.g., the primary lobe of the beam
exhibits non-circularity due to the off-axis position of detectors on
the focal plane, diffraction around the edges of instrument leads to
side lobes of the beam, or due to finite response time of detectors the
scan may not correspond to the instrument rotating around its beam
axis leading to the effective beam response at any pointing direction
being sensitive to the scan strategy, etc..  Regardless of the
specific origin of non-circularity in the beam response function, the
key common point is the potentially measurable SI violation generated
in the observed CMB maps. It has been argued whether the NC
beam effect in WMAP provides a plausible explanation of the observed
non-zero BipoSH measurements~\cite{IW-LA-HE-NG,HLC}. This paper
presents the results from a research program to directly assess
the nature and amplitude of SI violation that can arise due to NC-beam
response within the BipoSH representation and its implication for the
WMAP-7 BipoSH `anomaly'.

In the absence of circular symmetry, NC-Beam functions can be most generally
expanded in BipoSH basis and coefficients of the expansion,
$B^{LM}_{l_1 l_2}$, are referred to as {\em beam-BipoSH} coefficients. An ideal,
circular symmetric beam ensures vanishing beam-BipoSH coefficients for all $L>0$.
Importantly, the beam-BipoSH coefficients at $L>0$ not only capture NC-beam shape
but also include the additional modulation arising from the specific
scan strategy that sets the orientation of the NC-beam at any pixel.
We show that every non-zero beam-BipoSH coefficient, $B^{LM}_{l_1
l_2}$, would generate a corresponding non-zero CMB BipoSH coefficient,
$A^{LM}_{l_1 l_2}$, in the observed CMB map.

The analytic formulation presented is valid for any arbitrary beam
shape but progress to explicit expressions is possible within an
idealized `parallel-transport' (PT) scan strategy
approximation~\cite{TS-BR}. Further, mild deviations from circular
symmetry, that permit a perturbative approach, and residual, discrete
even-fold azimuthal, symmetry in the NC beam imply that the analytic
results at the leading order quadrupolar-non-circularity ($m=2$) captures most
of the significant features in the present observations. At the leading order of our
approximations, we provide simple explicit analytic expressions only
for BipoSH coefficients due to {\em mildly} NC-beams that retain {\em
discrete even-fold azimuthal and reflection symmetry} relevant to
understanding the WMAP-7 BipoSH results. Assumption of a reflection symmetric
beam function restricts the set to even parity BipoSH
coefficients~\footnote{Odd parity BipoSH coefficients are a measure of
the breakdown of reflection symmetry in instrumental
beams. Expressions for the odd parity BipoSH coefficients generated in
absence of reflection-symmetry of the beam are provided in the
appendix for completeness.}.  The published WMAP BipoSH results claim
that in ecliptic coordinates BipoSH at $(L=2,M=0)$ are the non-zero
BipoSH coefficients measured. Our general analytic formulation then suggests that a
PT-scan~\cite{TS-BR}) approximation should be a fairly good
simplifying approximation for the WMAP scan for the dominant $m=2$
mode of NC-beams in the ecliptic coordinates. We support this
assertion with maps quantifying the departures from this approximation
for a realistic WMAP scan. Further, the multiple hits of the beam with
varying orientation at any pixel reduces (average out)
the level of non-circularity. In particular, it can be
estimated in the specific case of WMAP scan, the $m=2$ mode, $b_{l2}$,
in the raw NC-beam can be expected to reduce by about a factor of
$\sim 0.45$, and higher $m$ mode maps are expected to suffer even
progressively greater reduction [see Fig.(\ref{fig:beamavghits})].

Assuming that the entire BipoSH signal measured by WMAP arises due to
NC-beam effect, we determine spectrum of quadrupolar ($m=2$) mode,
$b^{\rm eff}_{l2}$, of an effective beam (incorporating multiple visits
by the beam to sky pixels) that fits the measured BipoSH spectra. We
fit for a constant rescaling of the raw beam $b^{\rm eff}_{l2}= \alpha
b_{l2}$. Although, the best-fit $\alpha\sim 0.45$ obtained matches our
expectation, the reduced $\chi^2$ of the fit is unsatisfactory, and
would, at best, explain part of the BipoSH signal. We find that a
linear $l$ dependent correction to a constant scaling, $b^{\rm
eff}_{l2}= (\alpha + \beta \, l) \,  b_{l2}$ leads to a good simple
`phenomenological' fit to the measured WMAP-7 BipoSH spectra.

Detailed, computationally
expensive numerical simulations incorporating the two beam
differencing, real scan and map making may justify the linear
correction to the constant scaling, or, lead to the conclusion that
the entire BipoSH signal cannot be attributed to NC-beam leaving room
for other uncorrected systematic effects, or even hint at cosmological
SI violation.

We verify all analytic results and estimate error-bars through
extensive numerical simulations of SI CMB maps scanned by real space
convolution with corresponding NC-beam maps. The numerical simulations
presented here are limited to the PT-scan, single side, NC-beam
studied analytically. As explained later, without simplifying
assumptions, realistic simulations using the WMAP beam differencing
and scan are more compute intensive primarily because we find that the
results are fairly sensitive to side lobe and other spread-out
features (out to $\sim 10$ times the FWHM from the beam center),
present in the beam maps released by the WMAP
team~\cite{LAMBDA}. Hence, we defer the results from ongoing realistic
simulations to a future publication.

The paper is arranged as follows. Sec. \ref{bipfor} provides a brief
primer to the BipoSH formalism to characterize SI violations for keeping
the paper self contained. In Sec. \ref{ncbeambposh} we present a novel
expansion of the beam response function in the BipoSH basis, referred
to as {\em beam-BipoSH} coefficients in the article.  In
Sec. \ref{bipforbeam} we derive expressions for the CMB BipoSH
coefficients arising from the convolution of SI CMB anisotropy signal
with general NC-beams. We also provide the simpler explicit analytic
expressions for beam-BipoSH coefficients for a mildly NC
beam within the PT-scan approximation.  As a test case, we study the
Elliptical Gaussian (EG) NC-beam model that has minimal parameters with
clear interpretation. The beam spherical harmonic transforms ({\em
beam-SH}) have analytic expressions and permit simple, well controlled
perturbative treatment of the NC-effects. We also use this case to
test and qualify our numerical simulations against analytic
results. In Sec.~\ref{BipoSHWMAP}, we get closer to addressing the
WMAP-7 BipoSH measurements by first computing the BipoSH spectra based
on the beam-SH of single-side, raw beam maps
(Sec.~\ref{BipoSHWMAPrawbeam}). In Sec.~\ref{BipoSHbl2fit} we fit for
parametrized scaling functions, $f_l$ of the leading order raw beam-SH
that would reproduce the WMAP-7 BipoSH spectra measurements as arising
entirely from uncorrected NC beam effect.  Sec.~\ref{conclusions} has
discussions and presents the conclusions of this paper. Detailed steps
of all analytical calculations are provided for completeness in
Appendix~\ref{app:beam-biposh} and
\ref{app:cmb-biposh}. Appendix~\ref{beamavghits} provides an analytic
derivation of an effective beam arising from multiple hits at varying
orientations in terms of a rescaled beam-SH the raw beam map.

\section{Primer: Bipolar Spherical Harmonic (${\rm  BipoSH}$) representation}\label{bipfor}

The CMB anisotropy signal is a random field on the surface of a sphere
and can, hence, be expanded in the spherical harmonic (SH) basis,
\begin{eqnarray}
\Delta T(\hat n)=\sum_{lm}a_{lm}Y_{lm}(\hat n)\,,
\end{eqnarray}
and equivalently represented in the set of random variables $a_{lm}$.
Statistical properties of a random field are characterized by its
\textit{n}-point correlation function.  SI implies
expectation values of all \textit{N}-point correlation functions are
invariant under the rotations of the sky. The statistical distribution
of fluctuations in temperature field are observationally consistent
with Gaussian statistics.  Hence, the two-point function $C(\hat n,
\hat n')\equiv \langle \Delta T(\hat n)\Delta T(\hat n')\rangle$ is
expected to completely encode all information pertaining to the CMB
anisotropy, $\Delta T(\hat n)$. For CMB sky respecting statistically
isotropy (SI), the correlation function only depends on the angle
between the two directions $\hat n_1$ and $\hat {n_2}$ and the two
point function can then be expanded in Legendre polynomials, as
\begin{eqnarray}
C(\hat n_1, \hat n_2)\equiv C(\hat n_1 \cdot \hat
n_2)=\sum^{\infty}_{l=0}\frac{(2l+1)}{4\pi}C_l P_{l}(\hat n_1 \cdot \hat n_2),
\end{eqnarray}
where the coefficients of expansion $C_l$ define the angular power
spectrum. The random spherical harmonic moments, $a_{lm}$ are then
statistically independent leading to a diagonal covariance matrix in
the SH basis,
\begin{eqnarray}
\langle a_{lm} a^{*}_{l'm'}\rangle=C_{l}\delta_{ll'}\delta_{mm'}
\label{SI_SH_cov}
\end{eqnarray}
which also implies that (the $m$-independent), $C_l$ encodes all the
information in a Gaussian, SI field on the full sky (complete sphere,
$\textbf{S}^{2}$). In general, SI violation leads to non-zero
off-diagonal elements (for specific subset of SI violations, the SH
covariance matrix remains diagonal, but $m$ dependent). The two point
correlation function, most generally, then depends on both the
directions $\hat n_1$ and $\hat {n_2}$ and not just on the angle
between them. The most convenient expansion basis for $C(\hat n_1,
\hat n_2)\not\equiv C(\hat n_1 \cdot \hat n_2)$ is the Bipolar
Spherical Harmonic(BipoSH)
basis\cite{AH-TS-03,AH-TS-04,AH-TS-NC,AH-TS-05,SB-AH-TS,AH-TS-06}.
 \begin{equation}\label{eq:BPOSH}
C(\hat{n}_{1},\hat{n}_{2}) =
\sum_{l_{1},l_{2},L,M}A_{l_{1}l_{2}}^{L
M}\{Y_{l_{1}}(\hat{n}_{1})\otimes Y_{l_{2}}(\hat{n}_{2})\}_{L M},
\end{equation}
where $A_{l_{1}l_{2}}^{L M}$ are BipoSH coefficients and the bipolar
spherical harmonic (BiPoSH) functions,
\begin{eqnarray}\label{eq:BipoSH}
&&\{Y_{l_{1}}(\hat{n}_{1})\otimes Y_{l_{2}}(\hat{n}_{2})\}_{L M}  \ = \nonumber\\
&& \qquad \sum_{m_{1}m_{2}} C_{l_{1}m_{1}l_{2}m_{2}}^{LM}Y_{l_{1} m_{1}}(\hat{n}_{1})\;
Y_{l_{2}m_{2}}(\hat{n}_{2}),
\end{eqnarray}
are irreducible tensor product of two spherical harmonics spaces that
form an orthonormal basis on $\textbf{S}^{2} \times \textbf{S}^{2}$
and, here, $C_{l_{1}m_{1}l_{2}m_{2}}^{LM}$ are Clebsch-Gordon
coefficients. The indexes of these coefficients satisfy the
triangularity conditions $|l_1-l_2|\le L\le l_1+l_2$ and $m_1+m_2=M$.
The transformation properties of BipoSH under rotations are similar to
spherical harmonics~\cite{varshalovich}.

The BipoSH coefficients can be shown to be linear combinations of
off-diagonal elements of the harmonic space covariance matrix,
\begin{eqnarray}\label{eq:gen-BipoSH}
A^{LM}_{l_1 l_2}=\sum_{m_1 m_2}\langle a_{l_1 m_1}a^{*}_{l_2
m_2}\rangle(-1)^{m_2}C^{LM}_{l_1 m_1 l_2 -m_2},
\end{eqnarray}
where $a_{lm}$'s are the spherical harmonic coefficients of the CMB
maps~\cite{AH-TS-03}. The SI part given in Eq.~(\ref{SI_SH_cov}),
corresponds to $(L=0, M=0)$, since , $C^{00}_{l m l' -m'} \propto
\delta_{ll'}\delta_{mm'}$.  {\em Non-zero BipoSH coefficients with
$L>0$ that capture SI violation also neatly encode the residual
symmetries of the two point correlation
function}~\cite{NJ-SJ-TS-AH2010}.  The correlation function is always
real and symmetric under the exchange of $\hat n_{1}$ and $\hat
n_{2}$, reflected in the following symmetry properties for BipoSH
coefficients,
\begin{eqnarray}
 A^{LM}_{l_1 l_2}\ &=&\ \left\{
\begin{array}{ll}
 (-1)^{l_{1}+l_{2}-L+M}A^{*L -M}_{l_1 l_2}\\ \\
(-1)^{l_{1}+l_{2}-L}A^{LM}_{l_2 l_1}\\
\end{array}
\right. \,.
\end{eqnarray}
As a consequence, BipoSH coefficients are always real if $M=0$ and
$l_1+l_2+L$ is even and always imaginary if $M=0$ and $l_1+l_2+L$ is
odd.  If the observed CMB sky is statistically isotropic, it can be
shown that all the BipoSH coefficients vanish except the coefficients
of the form $A^{00}_{l l}$ which can be expressed in terms of the CMB
angular power spectrum~\cite{AH-TS-03},
\begin{eqnarray}\label{eq:statiso}
A^{LM}_{l_1 l_2}=(-1)^{l_1} C_{l_1}\prod_{l_1}\delta_{l_1 l_2}\delta_{L0}\delta_{M0},
\end{eqnarray}
where, for brevity, the standard notation $\prod_{a
b...c}=\sqrt{(2a+1)(2b+1)...(2c+1)}$ is introduced and employed in the
rest of this paper~\cite{varshalovich}.

Recently it has also been realized that BipoSH space representation
separates into two distinct classes,
\begin{itemize}
\item{} Even parity BipoSH coefficients $A^{LM^{(+)}}_{l_1 l_2}$
where $l_1+l_2+L$ is even obey  $A^{LM^{(+)*}}_{l_1 l_2}= (-1)^M A^{L\, -M^{(+)}}_{l_1
l_2}$ , and 
\item{} Odd parity BipoSH coefficients $A^{LM^{(-)}}_{l_1 l_2}$ with
$l_1+l_2+L$ odd, obey $A^{LM^{(-)*}}_{l_1 l_2}= (-1)^{M+1} A^{L\,
-M^{(-)}}_{l_1 l_2}$,
\end{itemize} 
where the `parity' terminology here is based on the analogy with
parity transformation properties of ordinary spherical harmonic
moments.

It is often easy to anticipate, or determine, the parity property of a
possible cause of SI violation. This distinction then provides very
valuable clues to the origin of SI violations e.g., weak lensing due
to scalar (even parity) and tensor (odd parity)
perturbations~\cite{LB-MK-TS2012}, anisotropic primordial power
spectrum (even)~\cite{AP-MK}, temperature modulation
(even)~\cite{DH-AL}, primordial homogeneous magnetic fields
(even)~\cite{AH-PhD,MA-TS}. Importantly, in the context of origin of
WMAP BipoSH measurements from NC-beam effect, the absence of
significant odd parity BipoSH would justify the simple
parallel-transport (PT) scan approximation adopted in our analytic
approach.

\section{${\rm Beam-BipoSH}$: Non-circular beams in ${\rm BipoSH}$ representation}
\label{ncbeambposh}

It is very important to characterize the non-circularity of the beam
through appropriate measures. We show that NC-beam is well
represented in BipoSH space.  Beam response function characterizes the
angular dependence of the sensitivity of the apparatus around the
pointing direction $\hat n_{1}$.  We expand the beam function around
the pointing direction $\hat n_1$ in Spherical Harmonic (SH) basis as,
\begin{eqnarray}
B(\hat n_{1},\hat n_{2})=\sum_{lm}b_{lm}(\hat n_{1})Y_{lm}(\hat n_{2}).
\label{beamSHexp}
\end{eqnarray}
The beam-SH coefficients when the beam is pointed at $\hat n_1$, are given by 
\begin{eqnarray}
b_{lm}(\hat n_{1})=\int d\Omega_{\hat n_{2}}B(\hat n_{1},\hat n_{2})Y_{lm}(\hat n_{2}).
\label{beamSH}
\end{eqnarray}
The SH transform of beam at arbitrary pointing direction, $\hat
n\equiv(\theta,\phi)$ is given by {\em beam-SH}, $b_{l m'}(\hat z)$ --
the SH transform of the beam pointed along $\hat z$, the North
pole~\cite{TS-BR},
\begin{eqnarray}\label{eq:b2}
b_{lm}(\hat n)=\sum_{m'}b_{l m'}(\hat z) D^{l}_{m
m'}(\phi,\theta,\rho(\hat n)),
\label{beamSHz}
\end{eqnarray}
where Wigner D-functions $D^{l}_{mm'}(\alpha,\beta,\gamma)$ are the
matrix elements of the rotation operator ($0\leq\alpha<2\pi,\
0\leq\beta<\pi,\ 0\leq\gamma<2\pi$) and $\alpha, \beta, \gamma$ are
the Euler angles that rotate the $\hat z$-axis to the pointing
direction $\hat n=(\theta,\phi)$ and the angle $\rho(\hat n)$ specifies
the orientation of the NC-beam with respect to the local Cartesian
$(\hat x\equiv\hat\phi, \hat y\equiv\hat\theta)$ aligned with
spherical coordinates~\cite{TS-BR}. Such a rotation can be realized
by fixing a coordinate system and performing anti-clockwise rotations,
first rotating about the $\hat z$-axis by an angle $\alpha=\phi$, then
rotating about new $\hat y$-axis by an angle $\beta=\theta$, and
finally about the new $\hat z$-axis by $\gamma=\rho(\hat n)$.
Rotation is such that the beam pointing direction, $\hat n$ and
$\rho(\hat n)$ is the relative orientation of the beam around its axis
set by the the scan pattern of the experiment.

In this work we introduce a more general NC-beam characterization in
the BipoSH representation. Since, a general NC-Beam function depends
on two vector directions, it can be expanded in the BipoSH basis (see
Sec.~\ref{bipfor}),
\begin{eqnarray}
&&B(\hat n_{1},\hat n_{2}) \ = \label{eq:beambipolar}\\
&& \qquad \sum_{l_1 l_2 L M} B^{LM}_{l_1 l_2}
\sum_{m_1 m_2} C^{LM}_{l_1 m_1 l_2 m_2} Y_{l_1 m_1}(\hat n_{1}) Y_{l_2
m_2}(\hat n_{2}), \nonumber
\end{eqnarray}
where the coefficients of expansion $B^{LM}_{l_1 l_2}$ are referred to
as {\em beam-BipoSH} coefficients . As will be
clear in the sections ahead, beam-BipoSH coefficients have the
advantage of globally capturing the additional effect of scan strategy together
with non-circularity of the beam.

The beam-BipoSH coefficient can be readily related to beam function
in spherical harmonic basis ,
\begin{equation}\label{eq:b1}
B^{LM}_{l_1 l_2}=\sum_{m_1 m_2}C^{LM}_{l_1 m_1 l_2 m_2} \int
d\Omega_{\hat n} b_{l_2 m_2}({\hat n})Y^{*}_{l_1 m_1}(\hat n)\,.
\end{equation}
A circularly symmetric beam function around the pointing direction can
be expanded in Legendre polynomials, $B(\hat n_{1},\hat n_{2})\equiv B(\hat
n_{1}\cdot\hat n_{2})={(4\pi)}^{-1}\sum_l (2l+1)B_{l}P_{l}(\hat
n_{1}\cdot\hat n_{2})$.  Inverse transforming
Eq.~(\ref{eq:beambipolar}) and using orthogonality of
BipoSH\cite{varshalovich}, we obtain beam-BipoSH coefficients for
circularly symmetric beam function,
\begin{eqnarray}\label{beamcircular}
B^{LM}_{l_1 l_2}=(-1)^{l_1}B_{l_1}\prod_{l_1}\delta_{l_1 l_2}\delta_{L0}\delta_{M0},
\end{eqnarray}
where $B_{l}$ is the commonly used Legendre transform of the beam
function in the circularized beam approximation.

\subsection{General expression for an arbitrary scan strategy}

Using Eq.~(\ref{eq:b2}) and (\ref{eq:b1}), it turns out that for any
arbitrary scanning strategy, the beam-BipoSH can be expressed in terms
of the beam-SH as,
\begin{eqnarray}\label{eq:beam-BipoSH}
&&B^{LM}_{l_1 l_2} \ =\ \sum_{m_1 m_2}C^{LM}_{l_1 m_1 l_2 m_2}\sum_{m'}b_{l_2
m'}(\hat z) \ \times \label{eq:genbeambiposh}\\
&&\quad \int^{\pi}_{\theta=0}\int^{2\pi}_0 d\phi D^{l_2}_{m_2
m'}(\phi,\theta,\rho(\theta,\phi))Y^{*}_{l_1
m_1}(\theta,\phi)\sin\theta d\theta. \nonumber
\end{eqnarray}
To separate the azimuthal ($\phi$) and polar ($\theta$) dependencies,
it is convenient to expressed Wigner-\textit{D} functions in terms of
Wigner-\textit{d} through following relation,
\begin{eqnarray}
D^{l}_{m m'}(\phi,\theta,\rho)={\rm e}^{-i m \phi}\,d^{l}_{m
m'}(\theta)\, {\rm e}^{-i m' \rho}.
\end{eqnarray}

Eq.~(\ref{eq:genbeambiposh}) is the most general expression of
beam-BipoSH coefficients for any given NC-beam specified through
$b_{lm}(\hat z)$ and scan pattern, defined by $\rho(\theta,\phi)$, in
any spherical polar coordinate system (e.g. ecliptic, galactic, etc.). The beam-BipoSH coefficients
can be evaluated numerically from the beam maps of the NC-beam.  When
the beam is circularly symmetric, $b_{lm}(\hat
z)=B_{l}\sqrt{\frac{2l+1}{4\pi}}\delta_{m0}$, and it is
straightforward to establish that Eq.~(\ref{eq:beam-BipoSH}) reduces to
Eq.~(\ref{beamcircular}).  Beam-BipoSH depend not only on NC-beam
harmonics but also on the scan-strategy that defines $\rho(\hat n)$.
We motivate a particular idealized scan pattern where $\rho(\hat n)$
is constant in a some coordinate system, that not only allows a
completely analytic treatment but is also reasonably well-justified in
the context of the BipoSH signature measured in WMAP-7.  Analytic
progress is less tedious when the beam has mild deviations from
circularity allowing a perturbative approach that retains only the
leading order terms. We validate numerically that these approximations
used are indeed fairly adequate for the explanation of the WMAP BipoSH
measurements.

\subsection{`Parallel-transport' scan approximation}

\begin{figure}[h]
\includegraphics[width=0.45\textwidth]{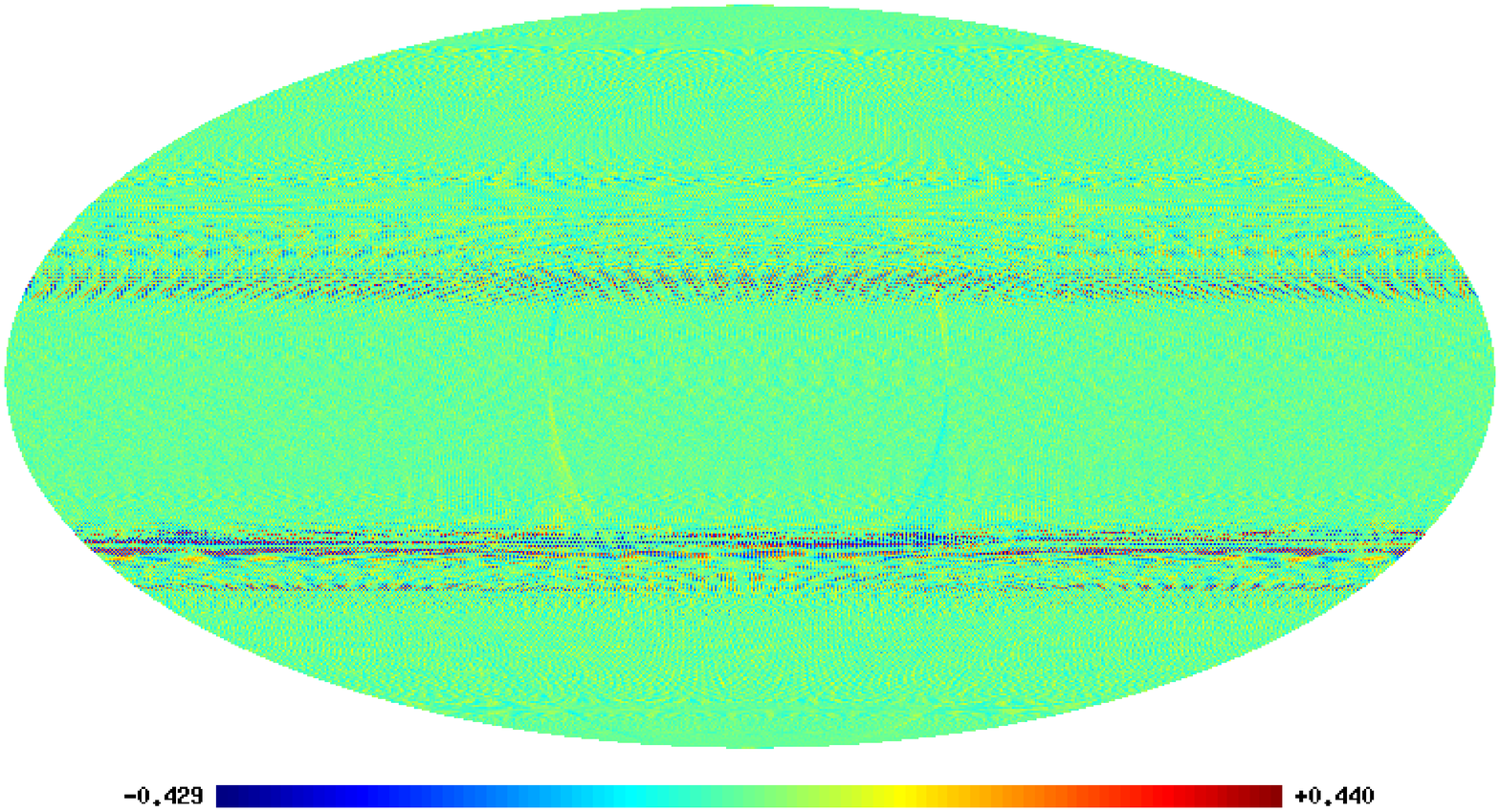} 
\includegraphics[width=0.45\textwidth]{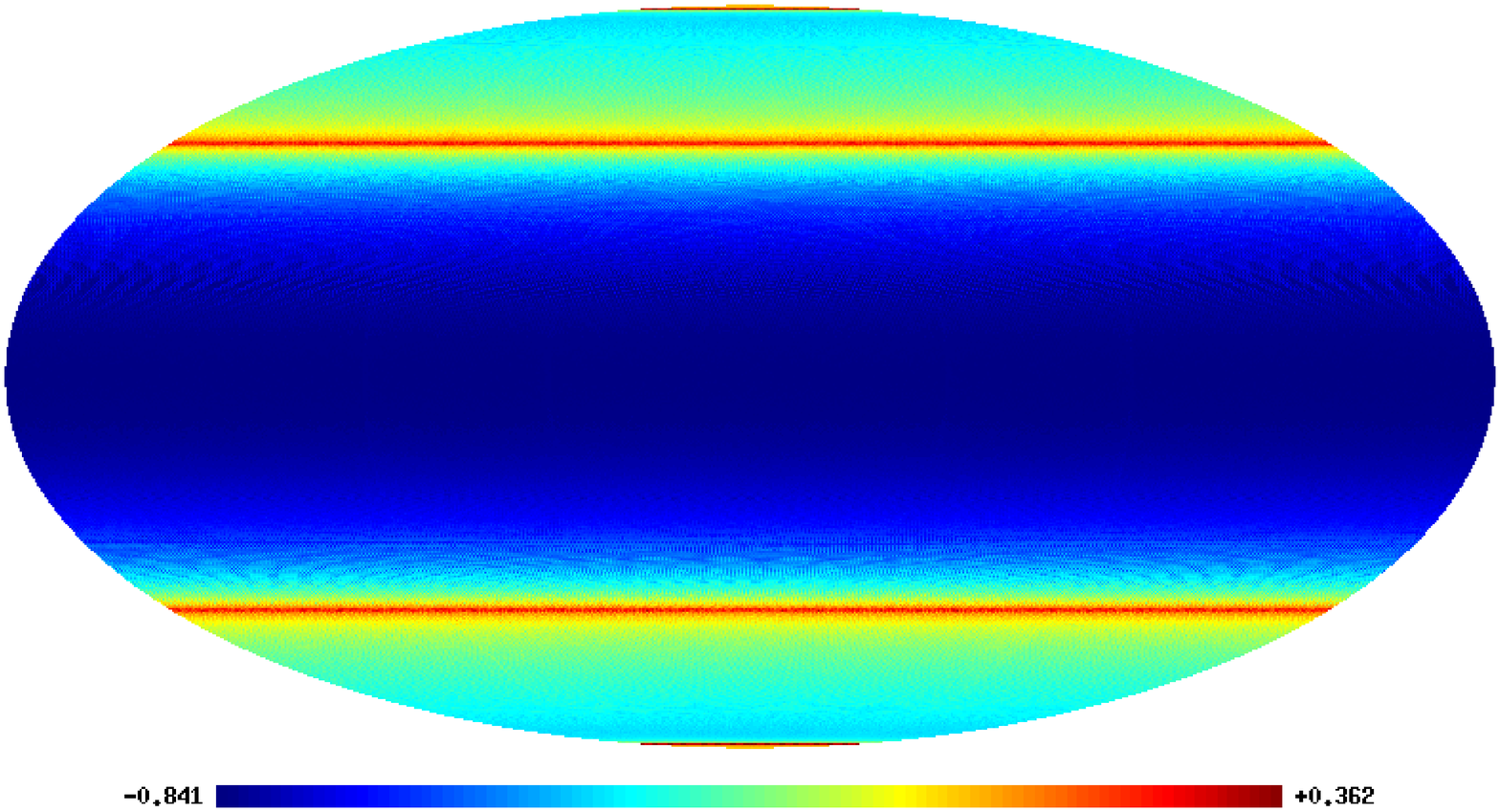} 
\caption{{\em Top:} Map of $\bar \rho(\hat n) = [N({\hat
n})]^{-1}\sum_i \rho_i(\hat n)$ defining the orientation of the beam
averaged over multiple hits, $i=1$ to $N({\hat n})$ for numerically
simulated one year of WMAP scan. At most pixels, $\bar \rho(\hat n)$
averages to a value close to zero. However, it also shows a very
intricate fine patterns of small regions with marked deviation from
zero.\newline {\em Bottom:} Map of $\langle \cos(2\rho) \rangle$ in ecliptic
coordinates for full year WMAP scan provides a rough guide to the
validity of the PT-scan approximation for the dominant $m=2$
non-circularity for the WMAP scan strategy. Note that over a wide band
of latitudes (blue region), the PT-scan approximation is roughly
valid. Strong deviation are seen at and beyond the two
latitude-symmetric thin bands (red) at moderately high latitudes in
North and South hemispheres.}
\label{fig:cos2rho}
\end{figure}

The general beam-BipoSH in Eq.~(\ref{eq:genbeambiposh}) can be tackled
analytically if when the scan pattern is such that $\rho(\hat n)$ is a
constant. {\em We refer to this case as `parallel-transport' (PT) scan
following~\cite{TS-BR}.}  The PT-scan approximation implies that the
orientation of the beam relative to the local longitude is constant at
any point on the sky.  A beam response function with reflection
symmetry will generate only real, non-vanishing beam-BipoSH
coefficients. The leading order $m=2$ dominates for a mildly
NC-beam. For real coefficients only the $\cos(2\rho)$ part is
important (see Eq.~(\ref{eq:beam-BipoSH})).  Fig.~\ref{fig:cos2rho}
shows the map for $\cos(2\rho)$ for a full year scan. For WMAP scan
pattern, the constant $\rho$ approximation for $m=2$ holds good over
the significant (blue) band around the equator in an ecliptic
coordinate system. Therefore, for a fair fraction of the sky part of
the map our constant $\rho$ approximation for $m=2$ mode of
non-circularity is a fair assumption. The compute-intensive numerical
comparison of results from this approximation with the real WMAP scan
is underway and will be reported in the near future.

For constant $\rho(\hat n)\equiv\rho$, the orthogonality relation,
\begin{eqnarray}
\int^{2\pi}_0 d\phi\exp(-i (m_1+m_2) \phi) \ = \ 2\pi\,\delta_{m_1,-m_2}\,,
\end{eqnarray}
implies that the integral over $\phi$ in Eq.~({\ref{eq:beam-BipoSH}}),
will separate from the integral over $\theta$ and would restrict the
non-zero beam-BipoSH to $M=0$, 
\begin{eqnarray}\label{eq:intmt}
&&B^{LM}_{l_1
l_2} \ = \ 2\pi\delta_{M0}\sqrt{\frac{(2l_{1}+1)}{4\pi}}\sum_{m'}b_{l_2
m'}(\hat z)\exp({-i m'\rho}) \ \times \nonumber\\
&& \qquad \sum_{m_2}C^{L0}_{l_1 -m_2 l_2 m_2}
\int^{\pi}_{\theta=0} d^{l_2}_{m_2 m'}(\theta)d^{l_1}_{-m_2
0}(\theta)\sin\theta d\theta.
\end{eqnarray}
We define
\begin{equation}
I^{l_1 l_2}_{m_2,m'} \ = \ (-1)^{m_2}\sqrt{\frac{(2l_1 +1)}{4\pi}}\int d^{l_2}_{m_2 m'}(\theta)d^{l_1}_{m_2 0}(\theta)\sin\theta d\theta.
\end{equation}
Here we have used the symmetry property of Wigner-\textit{d}
functions, $d^{l}_{m m'}=(-1)^{m-m'}d^{l}_{-m-m'}$.  Beam-BipoSH
coefficients derived in Eq.~(\ref{eq:intmt}) thus take the following form in constant scan approximation, $\rho(\hat n)=\rho$,
\begin{eqnarray}\label{eq:BB}
B^{LM}_{l_1 l_2}=
\quad 2\pi\delta_{M0}\sum_{m'}b_{l_2 m'}(\hat z)\exp({-i m'
\rho})\times\nonumber\\ \sum_{m_2}C^{L0}_{l_1 -m_2 l_2 m_2}I^{l_1 l_2}_{m_2,m'}.
\end{eqnarray}
For a mildly NC beam, the summation can be truncated at a
low value of $m'$.  Further, NC-beams that retain discrete even-fold
azimuthal symmetry will have $b_{lm}(\hat z)=0$ for odd values of $m$.
In BipoSH space, the consequence of discrete even-fold azimuthal and
reflection symmetric NC-beam translates to restricting non-zero
beam-BipoSH to $M=0$ and $l_{1}+l_{2}=\textrm{even}$. Imposing
reflection symmetry, beam-BipoSH are restricted to even values of
multipole $L$.  We derive explicit expressions for beam-BipoSH
coefficients up to leading order ($m'=2$) term .

To make analytical progress, we need to evaluate $I^{l_1
l_2}_{m_2,m'}$ for $m'=-2,0,2$.

It is important to note that, $m'=0$ corresponds to the usual circular
symmetric beam-BipoSH ($L=0$) coefficient. The non-circular $m'=\pm2$
will give rise to non-trivial (quadrupolar, $L=2$) beam-BipoSH. We derive beam-BipoSH coefficients for
circularly symmetric part of beam ($m'=0$) using Eq.~(\ref{eq:BB}).
For $m'=0$, $I^{l_1 l_2}_{m_2,m'}$ can be simplified to,
\begin{eqnarray}
I^{l_1 l_2}_{m_2,0}&=&(-1)^{m_2}\sqrt{\frac{(2l_1
+1)}{4\pi}}\frac{2}{2l_2+1}\delta_{l_1 l_2}.
\end{eqnarray}
Therefore, beam-BipoSH coefficients for circular part of the beam are of following form (refer
Appendix\ref{app:beam-biposh}),
\begin{eqnarray}\label{eq:beam-biposh-C}
B^{LM}_{l_1 l_2}&=&(-1)^{l_2}b_{l_2 0}(\hat z)\sqrt{4\pi}\delta_{l_1
l_2}\delta_{L0}\delta_{M0}\delta_{m'0}.
\end{eqnarray}
For $m'=\pm2$, the integrals are evaluated separately for the $m_2=0$
and $m_2\neq 0$ parts of the summation.
In the former case when $m_{2}=0$ and $m'=\pm2$, the integral 
\begin{eqnarray}
&&I^{l_1 l_2}_{0,\pm2}\ = \ \sqrt{\frac{(2l_1 +1)}{4\pi}} \ \times \nonumber\\
&&\quad\left\{
\begin{array}{ll}
0 & \mbox{if ($l_1 +l_2\equiv \textrm{odd}$)} \\ \\
0 & \mbox{if ($l_1 >l_2$)} \\ \\
4\sqrt{\frac{(l_{2}-2)!}{(l_{2}+2)!}} & \mbox{if ($l_1 <l_2$)} \\ \\
\sqrt{\frac{(l_{2}-2)!}{(l_{2}+2)!}}\big[\frac{4l_2}{(2l_2 +1)}-\frac{2l_2(l_2 +1)}{(2l_2 +1)}\big] & \mbox{if ($l_1 = l_2$)} \,.
\end{array}
\right.
\end{eqnarray}
  For $m_2\neq0$, $d^{l_2}_{m_2
\pm 2}(\theta)$ is recursively expanded in terms of $d^{l_2}_{m_2
0}(\theta)$ to evaluate $I^{l_1 l_2}_{m_2,\pm2}$ (refer Appendix
\ref{app:beam-biposh}).

This work is motivated by the highly significant measurements of
BipoSH published by the WMAP team~\cite{CB-RH-GH}. The
publication claimed that for the $L=2$ BipoSH spectra measured are
significantly non-zero only for $M=0$ in the ecliptic
coordinates. This has been independently confirmed by our BipoSH
measurements on WMAP-7, as well.  As shown in
Sec.~\ref{bipforbeam}, this implies that only beam-BipoSH with $M=0$
are relevant for understanding of the WMAP BipoSH results. Consequently,
it suggests that if NC-beam is responsible for non-zero BipoSH
measurements, the WMAP scan pattern is such that, in ecliptic
coordinate, $\rho(\hat n) \approx \textit{constant}$ could be a good
approximation for the dominant $m=2$ mode of the NC beams of
WMAP. This is being quantitatively assessed through more compute
intensive, numerical simulations with realistic WMAP scan patterns
(i.e., without invoking PT-scan approximation and incorporating
map-making from actual two side difference measurement). We take
advantage of the simplicity afforded in the PT-scan approximation to
proceed to explicit analytic expressions that readily provide broader
as well deeper insight than compute expensive numerical simulations.

PT-scan approximation also implies that only even-parity beam-BipoSH would be relevant
for our comparison to WMAP BipoSH results where the estimator used is
by definition strictly restricted to even-parity BipoSH. NC beam with reflection symmetry have non-vanishing beam-BipoSH with even-parity only,

\begin{eqnarray}
&&B^{LM}_{l_1 l_2}\equiv B^{LM^{(+)}}_{l_1
l_2}=2\pi\delta_{M0}\left(b_{l_2 2}(\hat z) {\rm
e}^{-i2\rho}+b^{*}_{l_2 2}(\hat z) {\rm
e}^{i2\rho}\right)  \nonumber\\
&&\qquad \times \ \left[C^{L0}_{l_1 0 l_2 0}I^{l_1
l_2}_{0,2}+\sum_{m_2\neq0}C^{L0}_{l_1 -m_2 l_2 m_2}I^{l_1 l_2}_{m_2,2}
\right]\,.
\label{beamBiposhm2}
\end{eqnarray}
Refer to Appendix~\ref{app:beam-biposh} for details. Note that a
constant $\rho$ can be absorbed as phase factor in the redefinition of
the complex quantity $b_{lm}(\hat z)$ essentially resetting the
orientation of the beam when pointed at North pole. Hence, the beam-BipoSH due to the NC part of the beam
in the PT-scan approximation is of the following form,
\begin{eqnarray}\label{eq:epbeambposh}
&&B^{LM^{(+)}}_{l_1 l_2} \ =\ 2\pi\delta_{M0}(b_{l_2 2}(\hat z)+b^{*}_{l_2
2}(\hat z)) \ \times \nonumber\\
&&\qquad \big(C^{L0}_{l_1 0 l_2 0}I^{l_1
l_2}_{0,2} \ + \sum_{m_2\neq0}C^{L0}_{l_1 -m_2 l_2 m_2}I^{l_1
l_2}_{m_2,2}\ \big).
\end{eqnarray}

We emphasize that the above expression for beam-BipoSH coefficient
holds for the PT-scan approximation (with constant $\rho({\hat n})$)
for a NC-beam that has reflection symmetry. 
Beam-BipoSH coefficient is a sum of circular part as in Eq.~(\ref{eq:beam-biposh-C}) and NC part of the beam function as in Eq.~(\ref{eq:epbeambposh}) and can be expressed as,
\begin{eqnarray}
B^{LM}_{l_1 l_2} =(-1)^{l_2}b_{l_2 0}(\hat z)\sqrt{4\pi}\delta_{l_1
l_2}\delta_{L0}\delta_{M0}\delta_{m'0}+\nonumber\\ 2\pi\delta_{M0}(b_{l_2 2}(\hat z)+b^{*}_{l_2
2}(\hat z)) \ \times \nonumber\\
\qquad \big(C^{L0}_{l_1 0 l_2 0}I^{l_1
l_2}_{0,2} \ + \sum_{m_2\neq0}C^{L0}_{l_1 -m_2 l_2 m_2}I^{l_1
l_2}_{m_2,2}\ \big).
\end{eqnarray}

Although we have restricted explicit analytic results presented in the
text to reflection symmetric beam functions, in general, odd parity
beam BipoSH will be non-vanishing in absence of the above mentioned
symmetries. Appendix~\ref{app:beam-biposh} provides expressions for
odd-Parity beam-BipoSH, $B^{LM^{(-)}}_{l_1 l_2}$, that can be used as a
measure of breakdown of reflection symmetry in NC
beam\footnote{Departure from reflection symmetry in the beam of a
full-sky CMB experiment, if ignored, also causes leakage of power from
the $\sim 500$ times stronger CMB dipole signal into higher multipole,
most importantly, contaminating the CMB quadrupole moment of the
angular power spectrum. This has been studied and estimates on WMAP
beam maps indicates the effect of reflection breakdown symmetry is
expected to be small, but not negligible~\cite{SD-TS2012}.}. Note that
the BipoSH estimator~\cite{DH-AL}, that differ by a factor from
original definition of Hajian \& Souradeep~\cite{AH-TS-03,AH-TS-06},
used by the WMAP team cannot be extended to odd-parity BipoSH,
However, it is possible to devise BipoSH estimators that can measure
odd-parity BipoSH spectra while matching that employed by WMAP for
even-parity BipoSH spectra~\cite{MK-TS2010}.

\section{ Non-circular beam imprint on ${\rm BipoSH}$ of CMB maps}\label{bipforbeam}

The observed CMB map is a convolution of the true CMB anisotropy sky with the
instrumental beam.  Instrument beam response functions of all
experiments have deviations from circular symmetry at some level.
However, circular symmetric beam response function around the pointing
direction is often assumed to simplify the analysis of the the beam
effect. This assumption, though, does affect many stages of CMB data
analysis and should be measured and characterized.

In this section we show that CMB maps made with NC beam
exhibit SI violation in an otherwise SI
cosmological CMB signal.  BipoSH coefficients have proved to be
robust, model independent, measure of SI violation. Further, BipoSH
representation provides clear insight into the nature and extent of
the residual symmetry and parity~\cite{NJ-SJ-TS-AH2010,LB-MK-TS2012}
and helps pin down the most plausible origin of SI violation.

The measured CMB temperature fluctuations map, $\widetilde{\Delta
T}(\hat n)$ is the convolution of true underlying CMB temperature
fluctuations $\Delta T(\hat n)$ with the instrument beam,

\begin{eqnarray}
\widetilde{\Delta T}(\hat n_{1})=\int d{\Omega_{\hat n_{2}}}B(\hat
n_{1},\hat n_{2})\Delta T(\hat n_{2}).
\end{eqnarray}
where the beam response function $B(\hat n_{1},\hat n_{2})$ gives the
sensitivity of an experiment around the pointing direction.  The
observed two point correlation function is,
\begin{eqnarray}
&&\tilde C(\hat n_{1},\hat n_{2})\equiv\langle\widetilde{\Delta T}(\hat
n_{1})\widetilde{\Delta T}(\hat n_{2}) \rangle \ = \nonumber\\
&& \qquad \int d\Omega_{n}\int
d\Omega_{n'}C(\hat n',\hat n) B(\hat n_{1},\hat n') B(\hat n_{2},\hat
n),
\label{eq:obs-corr}
\end{eqnarray}
where $C(\hat n',\hat n)=\langle\Delta T(\hat n')\Delta T(\hat
n)\rangle$ is the underlying correlation function determined by
cosmology.  It is evident from Eq.~({\ref{eq:obs-corr}}), that SI
violation can occur either due to rotational invariance breakdown of the
true underlying temperature correlation function, $C(\hat n_{1},\hat
n_{2})\not\equiv C(\hat n_{1}\cdot \hat n_{2})$, or, due to the
breakdown of circularity in beam response function $B(\hat n_{1},\hat
n_{2})\not\equiv B(\hat n_{1}\cdot \hat n_{2})$, or, both.

If the underlying CMB signal respects SI
symmetry as widely assumed in cosmology, then
\begin{eqnarray}
C(\hat n_{1},\hat n_{2})=\sum_{l}\frac{2l+1}{4\pi}C_{l}\,\,W_{l}(\hat
n_{1},\hat n_{2}),
\end{eqnarray}
where $W_{l}(\hat n_{1},\hat n_{2})$ is the {\em elementary window
function}~\cite{TS-BR} that accounts for the effect of finite
resolution of beam function, given as
\begin{equation}
W_{l}(\hat n_{1},\hat n_{2})=\int d\Omega_{\hat n}\int d\Omega_{\hat
n'}B(\hat n_{1},\hat n)B(\hat n_{2},\hat n')P_{l}(\hat n\cdot \hat
n').
\end{equation}

Inverse transform of Eq.~(\ref{eq:BPOSH}), yields the CMB BipoSH
coefficients, $\tilde A^{LM}_{l_1 l_2}$ in terms of beam-BipoSH
coefficients as
\begin{eqnarray}\nonumber
 \tilde A^{LM}_{l_1 l_2}=\int d\Omega_{n_1}\int d\Omega_{n_2}\tilde
 C(\hat n_{1},\hat n_{2})\{Y_{l_{1}}(\hat{n}_{1})\otimes
 Y_{l_{2}}(\hat{n}_{2})\}^{*}_{L M}
\end{eqnarray}

Hence, when the underlying CMB signal is statistically isotropic and
beams are circular, BipoSH coefficient are expected to vanish for
$L>0$, refer Eq.~(\ref{eq:statiso}),
\begin{equation}
 \tilde A^{LM}_{l_1
 l_2}=(-1)^{l_1}\prod_{l_1}C_{l_1}B^{2}_{l_1}\delta_{l_1
 l_2}\delta_{L0}\delta_{M0}.
\end{equation}

In Sec.~\ref{ncbeambposh}, we showed that non-circularity of the beam
response function is captured by beam-BipoSH coefficients.  The BipoSH
coefficients of CMB map for a reflection symmetric beam function can then be derived in terms of beam-BipoSH
coefficients as (see Appendix~\ref{app:beam-biposh}),
\begin{eqnarray}
&&\tilde A^{LM}_{l_1 l_2} \ =\ \sum_{l}(-1)^{l} C_{l} \sum_{L_1 M_1 L_2 M_2}B^{L_1
M_1}_{l_1 l}B^{L_2 M_2}_{l_2 l} \ \times \nonumber\\
&&\qquad\prod_{L_1 L_2}C^{LM}_{L_1 M_1
L_2 M_2} {\begin{Bmatrix} l & l_2 & L_2 \\ L & L_1 & l_1
\end{Bmatrix}}\,.
\label{biposhbeam}
\end{eqnarray}
In Sec.~\ref{ncbeambposh}, we also argued that it is convenient, as
well as observationally motivated, to carry out the analytic analysis
in a coordinate system where parallel-transport (PT) scan
approximation holds and, consequently, the non-zero beam-BipoSH are
restricted to $M=0$. Eq.~(\ref{biposhbeam}) then dictates that the
corresponding BipoSH coefficients of the CMB maps are also restricted
to $M=0$ and are given by

\begin{eqnarray}
&&\tilde A^{L0}_{l_1 l_2}=\sum_{l}(-1)^{l}C_{l} \ \times \nonumber\\
&&\qquad \sum_{L_1 L_2}B^{L_1 0}_{l_1 l}B^{L_2 0}_{l_2 l} \prod_{L_1 L_2}C^{L0}_{L_1 0 L_2 0}
{\begin{Bmatrix}
l & l_1 & L_2 \\
L & L_1 & l_2 
\end{Bmatrix}}\,.
\label{biposhbeamM0}
\end{eqnarray}
It turns out that due to triangularity condition ($|L_1-L_2|\leq L\leq L_1+L_2$),
the most dominant terms in the above summation are $\{L_1=L,
L_2=0\}\ \textrm{and}\ \{L_1=0, L_2=L\}$ as they are proportional to $B^{0 0}_{l_1 l}B^{L 0}_{l_2 l}$ and 
$B^{L 0}_{l_1 l}B^{0 0}_{l_2 l}$, which in turn depends on the product of SH coefficients $b_{l0}b_{l2}$.
In the beam response function $b_{l0}$ is significantly larger than $b_{l2}$, therefore the $b_{l0}b_{l2}$ term will be much larger than $b_{l2}b_{l2}$ which will contribute as second order terms in 
Eq.~(\ref{biposhbeamM0}).

The BipoSH estimator used by the WMAP team~\cite{CB-RH-GH,DH-AL},
differs from the original BipoSH definition in Hajian \&
Souradeep~\cite{AH-TS-03} by a factor of ${\prod_{L}}/({\prod_{l_1
l_2}C^{L0}_{l_1 0 l_2 0})}$ and are restricted to only even-parity
BipoSH. {\em We insert this factor in the BipoSH expression in order
to allow comparison to the published WMAP measurements more
transparent}~\footnote{Note that this factor in WMAP-BipoSH estimator
strictly restricts BipoSH considerations to the even parity sector
since $C^{L0}_{l0 l'0} =0$ for odd values of the sum $L=l+l'$. In the
context of NC-beams, this would be a handicap if reflection symmetry
is violated leading to odd-parity BipoSH coefficients (see
Sec. \ref{ncbeambposh}). Also it is blind to a number of other
interesting possibilities with odd-BipoSH signals. }

\begin{eqnarray}\label{biposh-constantscan}
\tilde A^{L0}_{l_1 l_2} \rightarrow \frac{\prod_{L}}{\prod_{l_1
l_2}C^{L0}_{l_1 0 l_2 0}} \tilde A^{L0}_{l_1 l_2}\,.
\label{Bmatrix}
\end{eqnarray}

SI violation signals in WMAP-7 were measured in two BipoSH spectra,
$A^{20}_{ll}$ and $A^{20}_{l-2 l}$, we provide explicit leading order
expressions for these coefficients arising from the NC-beam as,
\begin{eqnarray}
\label{eq:WMAP1}
\tilde A^{20}_{ll} & = &\frac{(-1)^{l}\,2\sqrt{5}\,C_{l}\,B^{00}_{ll}B^{20}_{ll}}{(\prod_l)^{3} C^{20}_{l
0 l 0}}\, ,\\
\tilde A^{20}_{l-2 l} &=&\frac{\sqrt{5}(-1)^{l}}{\prod_{{l-2}
l}C^{20}_{{l-2} 0 l 0}} \ \times \nonumber\\
&&\  \Big[\frac{C_{l-2}B^{00}_{{l-2} {l-2}}B^{20}_{l
{l-2}}}{\prod_{l-2}}+ \frac{C_{l}B^{00}_{l l}B^{20}_{{l-2}
l}}{\prod_{l}} \Big]\,.\label{eq:WMAP2}
\end{eqnarray}

Note, that BipoSH expression in Eqs.~(\ref{eq:WMAP1}) and
(\ref{eq:WMAP2}) are provided in the scaled form that matches the
BipoSH estimator employed by the WMAP team.

\subsection{ BipoSH from Elliptical Gaussian beam model}

Elliptical-Gaussian (EG) functions provide a simple model NC-beam
extension to the often used circular-symmetric Gaussian beam
function. The non-circularity is clearly parametrized by the
eccentricity of the elliptical iso-contour lines.  An EG-beam function
pointed along $\hat z$ axis can be expressed in spherical polar
coordinates, as
\begin{equation} 
B(\hat{z},\hat{n}) \ = \ \frac{1}{2\pi\sigma_1\sigma_2}\exp \left[ -\frac{{\theta^2}}{2\sigma^2(\phi)} \right],
\end{equation}
where the azimuth angle dependent beam-width $\sigma(\phi_1) \ \equiv
\ [\sigma_1^2/(1+\epsilon \sin^2\phi_1)]^{1/2}$ is given by Gaussian
widths $\sigma_1$ and $\sigma_2$ along the semi-major and semi-minor
axes. The non-circularity parameter $\epsilon \ = \
(\sigma_1^2/\sigma_2^2 - 1)$ is related to eccentricity $e =
\sqrt{1-\sigma_2^2/\sigma_1^2}$ of the elliptical iso-contours.  As
expected, the EG beam reduces to circular Gaussian
beam for zero eccentricity ($e=0$). Higher the value of eccentricity,
stronger the deviation from circularity.  An analytical expression for
the beam-SH of EG-beam  is available~\cite{TS-BR}: due to
reflection symmetry, for odd $m$, $b_{lm}(\hat z) = 0$, and for even $m$,
\begin{eqnarray}\label{eq:blm-EG}
&&b_{lm}(\hat z)\ \ = \ \ \sqrt{\frac{(2l+1)(l+m)!}{4\pi(l-m)!}}(l+\frac{1}{2})^{-m} \ \times \\
&& \quad I_{m/2}\Bigg[
(l+\frac{1}{2})^{2}\frac{\sigma^{2}_{1}e^{2}}{4}\Bigg]\exp\big[{-(l+\frac{1}{2})^{2}\frac{
\sigma^{2}_{1}}{2}({1-\frac{e^{2}}{2}})\big]} \, . \nonumber
\end{eqnarray}
where $I_\nu(x)$ is the modified Bessel function.  The
discrete even-fold azimuthal symmetry and reflection symmetry dictates
$b_{lm}=0$ for odd $m$.  The reality condition of beam,
$b^{*}_{lm}=b_{lm}$ for even $m$, then implies $b_{l -m}=b_{lm}$.
\begin{figure}[h]
\centering
\includegraphics[height=0.45\textwidth,angle=-90]{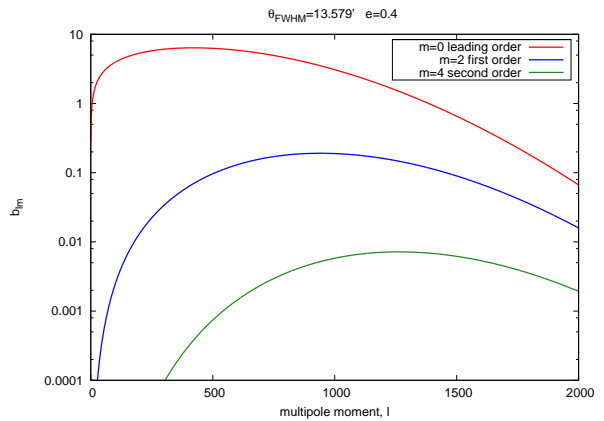}
\caption{Harmonic transforms of an EG beam with
$\theta_{\textit{FWHM}}=13.579'$ and ellipticity $e=0.4$. The circular
symmetric component of the NC EG-beam is given by is $m=0$ and $m=2$
captures the leading order EG-beam effect. }
\label{fig:ellgauss-blm}
\end{figure}

For Elliptical-Gaussian beams, the ratio $b_{lm}/b_{l0}$ dies down
rapidly with $|m|$. Hence, for mild eccentricities of the Elliptical
beam, it is sufficient to consider only the lowest $m=2$ mode.  In
Fig.~\ref{fig:ellgauss-blm}, we plot spherical harmonic transform of
an Elliptical-Gaussian beam with $\theta_{\textit{FWHM}}=13.579'$,
which is close to $\theta_{\textit{FWHM}}$ of W band of WMAP with
ellipticity $e=0.4$ that best matches WMAP-7 W-band BipoSH spectra
amplitude.  Similarly, we also consider a EG-beam with
$\theta_{\textit{FWHM}}=17.7036'$ close to the V-band beam. The best
fit to V-band BipoSH spectra corresponds ellipticity $e=0.46$. For
analysis purpose, it is a good enough approximation to restrict to
$m=0,\pm2$ modes.  We estimate beam-BipoSH coefficient in
Eq.~(\ref{eq:epbeambposh}) for the case of ``non-rotating'' beams,
$\rho=0$ by using the closed analytical form of $b_{lm}$'s as in Eq.~(\ref{eq:blm-EG}).  Finally,
using Eq.~(\ref{eq:WMAP1}) and Eq.~(\ref{eq:WMAP2}) we obtain the CMB
BipoSH spectra $A^{20}_{ll}$ and $A^{20}_{l-2 l}$ that were measured
to be non-zero in WMAP-7. We verify the analytical results and compute
the error bars on the BipoSH coefficients using statistically
isotropic Gaussian simulations convolved with EG-beam functions.
\begin{figure*}[!htbp]
\includegraphics[height=0.45\textwidth, angle=-90]{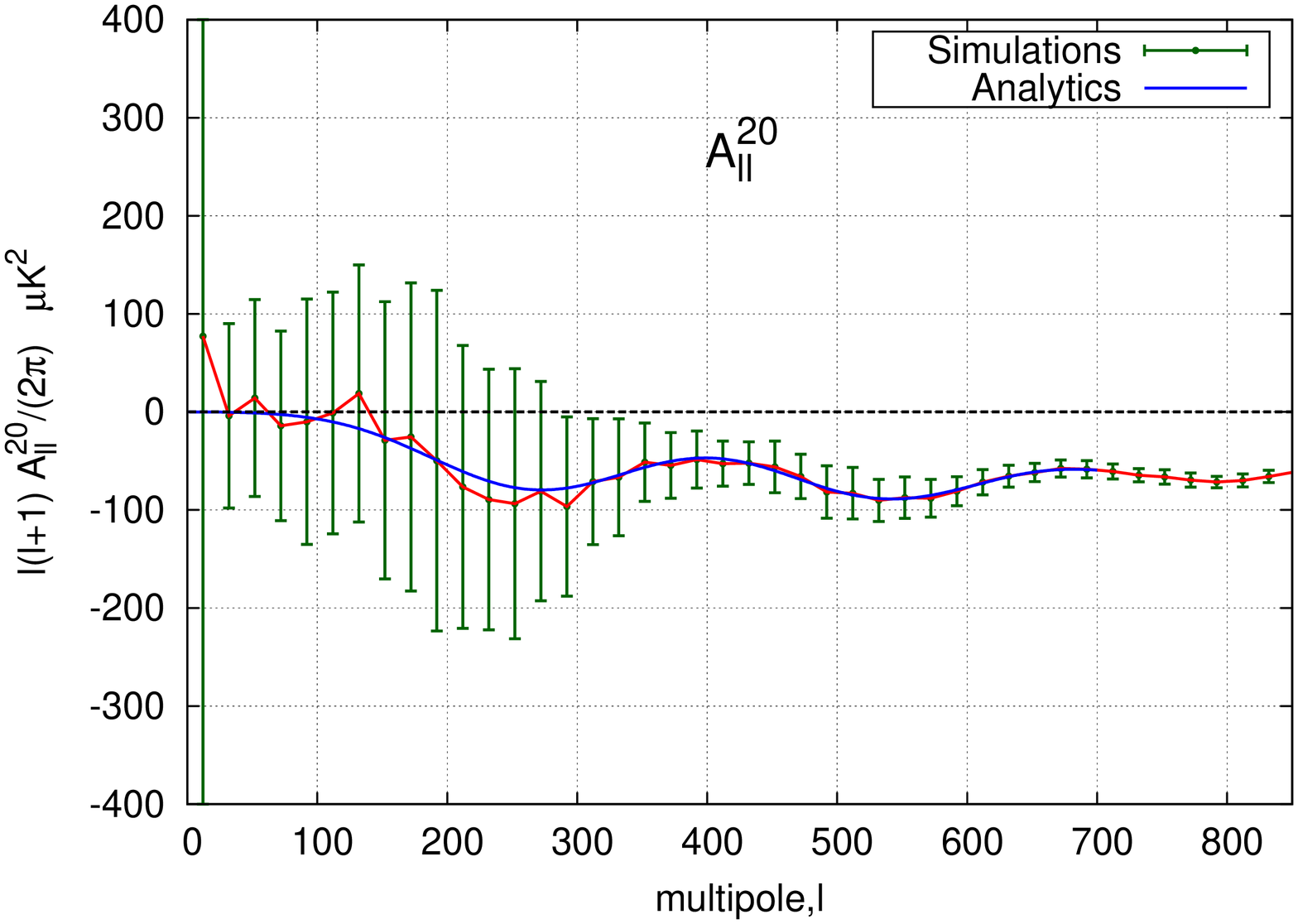}
\includegraphics[height=0.45\textwidth,angle=-90]{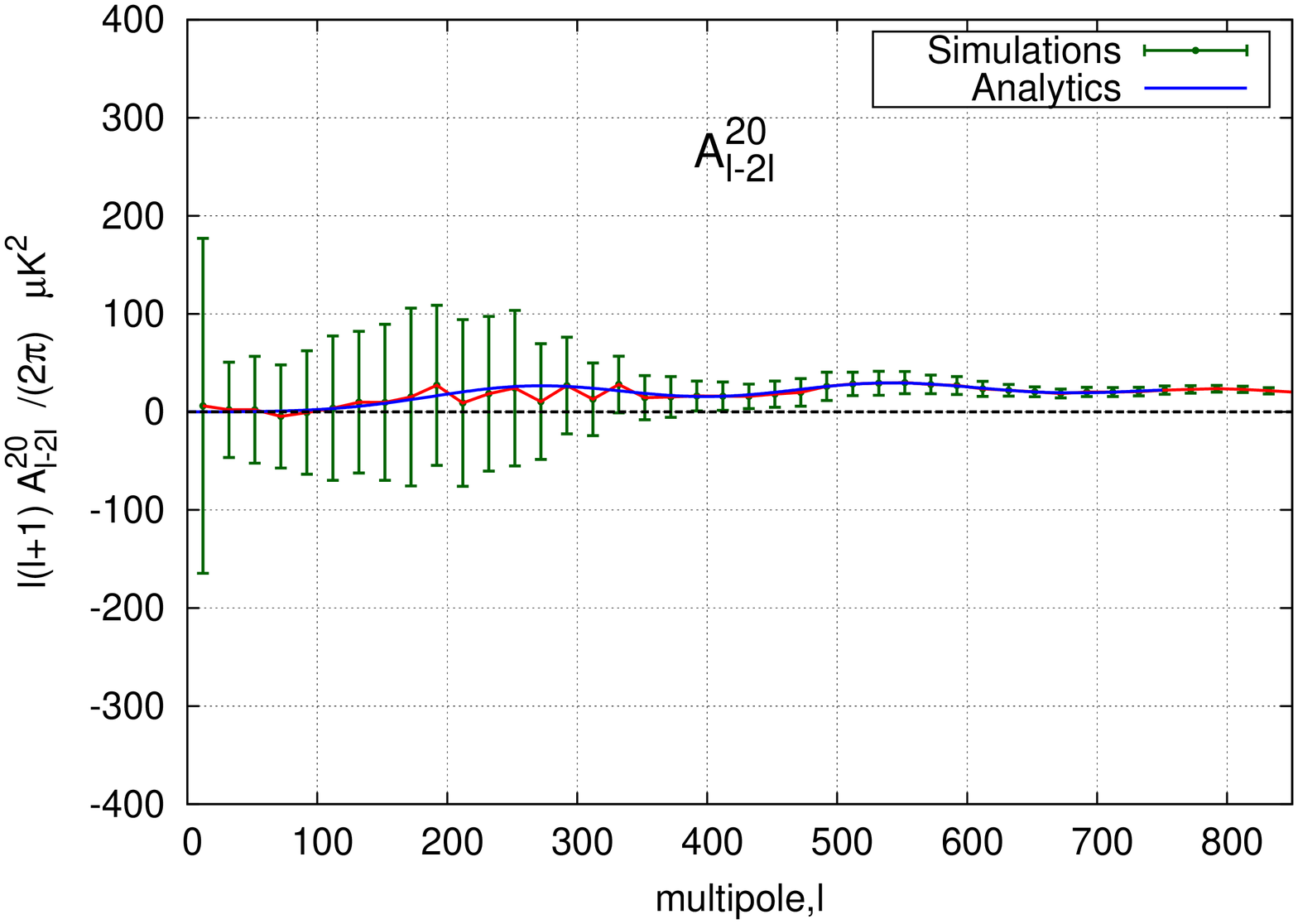}
\includegraphics[height=0.45\textwidth,angle=-90]{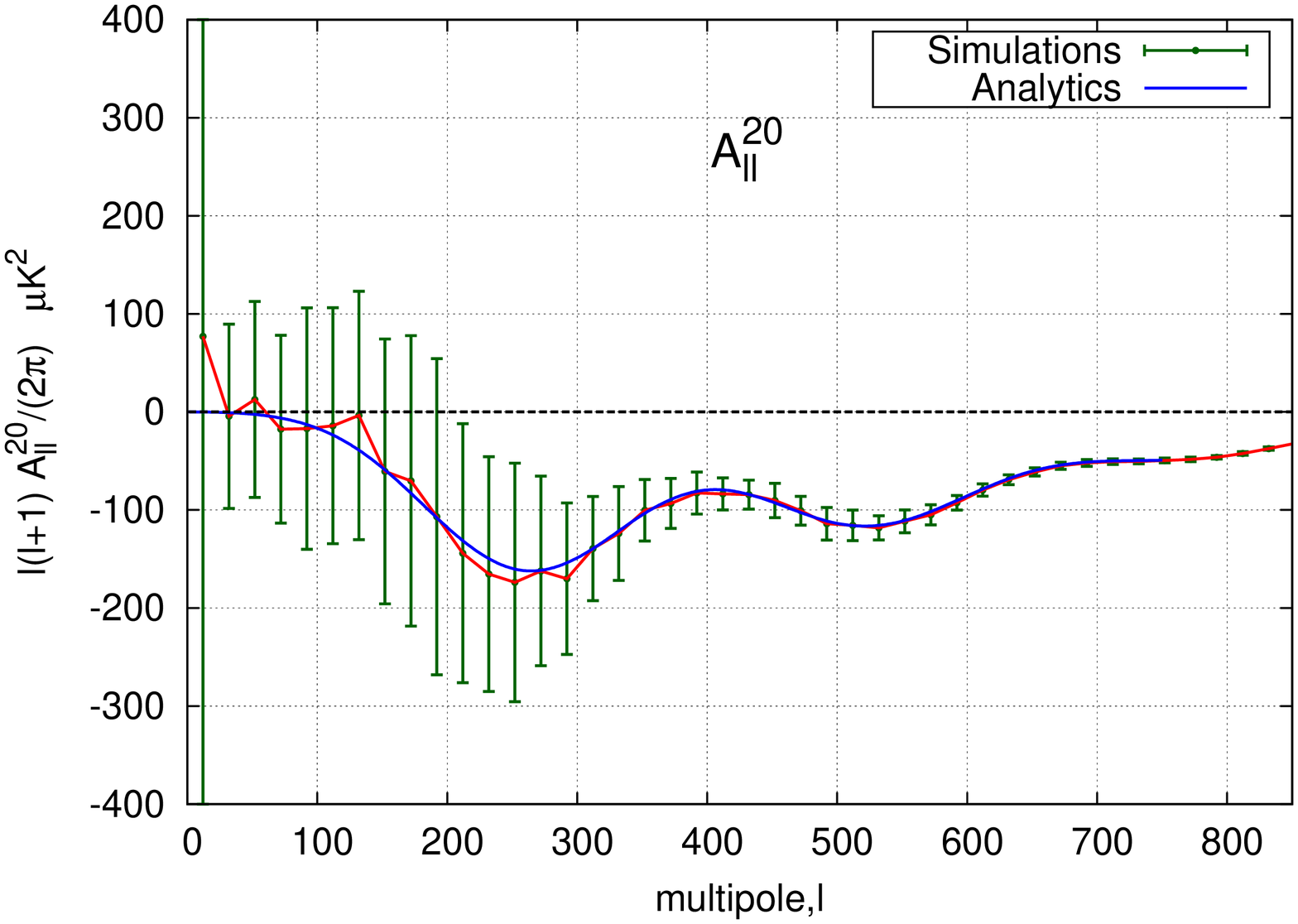}
\includegraphics[height=0.45\textwidth,angle=-90]{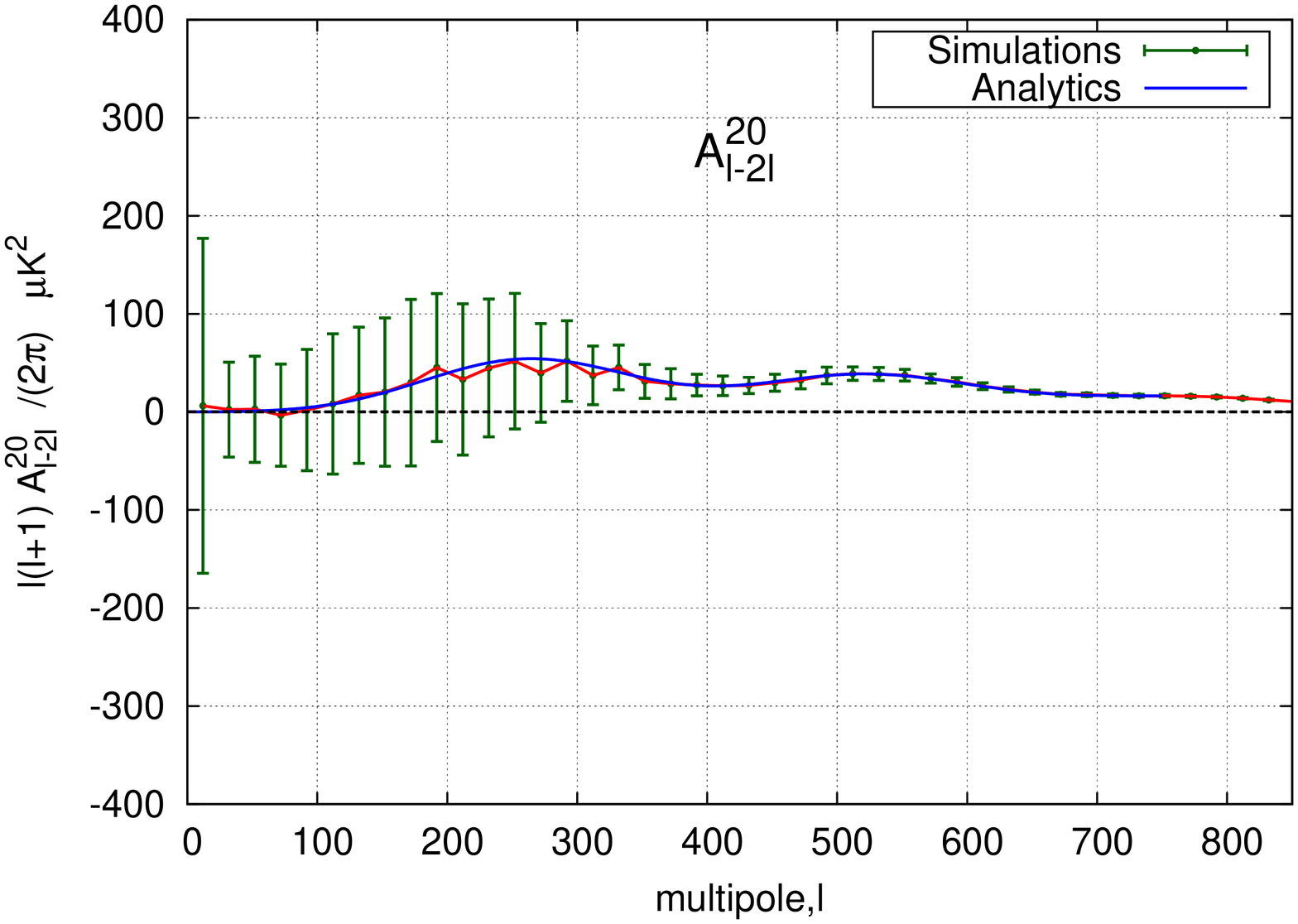}
	\caption{BipoSH spectra $A^{20}_{ll}$ and $A^{20}_{l-2 l}$ for
	 an EG beam with : ({\em Top})
	 $\theta_{\textit{FWHM}}=13.579'$ and ellipticity $e=0.4$;
	 ({\em Bottom}) $\theta_{\textit{FWHM}}=17.7036'$ and
	 ellipticity $e=0.46$ smooth (blue) curve are binned
	 CMB-BipoSH computed using the analytic expressions. The red
	 curve with corresponding error-bars are obtained from 100
	 numerical simulated SI maps convolved with the same
	 EG beam. The underlying SI signal
	 corresponds to the best fit $\Lambda$-CDM model from WMAP7
	 \label{fig:ellgauss-WV}}
\end{figure*}
In Fig.~\ref{fig:ellgauss-WV}, the analytic calculations of CMB-BipoSH
for two cases of EG-beam are overlaid on estimates from average
measurements with error-bars from 100 numerical simulations of noise
free SI maps convolved with corresponding NC-beams. 
BipoSH coefficients obtained from EG beams serve to crosscheck and validate our
analytical expression derived and puts a check on the numerical simulation of
CMB maps convolved (in real space) with an NC-beam.

It is clear that NC-beams generate detectable levels of non-zero CMB
BipoSH. The CMB-BipoSH spectra $A^{20}_{ll}$ and $A^{20}_{l-2 l}$ are
both proportional to the the leading order NC correction
$b_{l0}b_{l2}$, and the underlying SI CMB power spectrum (see
Eqs.~(\ref{biposhbeamM0}) and (\ref{beamBiposhm2})). It is instructive
to compare the EG-model beam predictions to the WMAP BipoSH
measurements.  Using Elliptical-Gaussian beams with FWHM values close
to the WMAP W and V band reproduces significant detectable peaks in the
BipoSH spectra. The amplitudes can be matched by choosing appropriate
values of effective eccentricity for the V ($e=0.46$) and W bands
($e=0.4$). Although, the eccentricity of the WMAP beams can be
estimated for raw beam maps~\cite{SM-AS-TS}, it is expected that
multiple visits by the beam at any pixel with varying orientation in
WMAP scan would tend to symmetrize the beam and reduce the effective eccentricity to smaller
values. The eccentricities that reproduce the correct amplitudes for
the CMB-BipoSH roughly correspond to the reduction also indicated from
comparing the corrections to the angular power
spectrum~\cite{SM-AS-SR-RS-TS}).  The fact that both the $A^{20}_{ll}$
and $A^{20}_{l-2 l}$ spectra, are roughly matched in amplitude and
qualitative peak structure with the correct relative sign, for
identical ellipticity parameters, then suggests that the WMAP-7 BipoSH
measurements may have actually captured subtle uncorrected NC-beam
effects. This claim is made strong through realistic simulations in the following sections.

\section{BipoSH signatures of  Non-Circularity in  WMAP beams }
\label{BipoSHWMAP}

While CMB BipoSH from the test EG beam motivates a
more careful estimation of the NC-beam effect in WMAP-7 maps, these
fail to reproduce certain features of WMAP BipoSH
measurements; viz. the change in sign of the V-band $A^{20}_{ll}$ at
large multipoles.  This, however, is not surprising.  It is
widely-known that the WMAP beams deviate from a Gaussian profile and
that an EG beam is not a good
approximation~\cite{Pag_WMAP03,SM-AS-TS,Hin_WMAP07}. In
section~\ref{BipoSHWMAPrawbeam}, we first use the SH transform of the
raw WMAP beam maps to compute the CMB-BipoSH under PT-scan
approximation.  Interestingly, we recover qualitative features in
BipoSH spectra measurements that were not captured by the Elliptical
Gaussian beams. As expected the amplitude of the CMB-BipoSH based raw
beam are much higher in amplitude relative to the measurements. A
rudimentary numerical superposition of raw beam transforms to mimic
the effect of multiple hits with varying orientation at a pixel (using
a WMAP scan), leads us to expect a constant scaling of the leading
order $b_{l2}$ by a factor $\alpha\sim 0.45$, as shown in Fig.~\ref{fig:beamavghits}. 
Moreover detailed analytical
treatment presented in the Appendix~\ref{beamavghits}, allows for
deviations from PT-scan approximation in averaging over multiple hits,
reveals multipole $l$ dependent correction, $f_l$ to the constant
scaling that depends on the detailed scan strategy of the CMB
experiment over all multiple hits.
\begin{figure}
\includegraphics[height=0.45\textwidth,angle=-90]{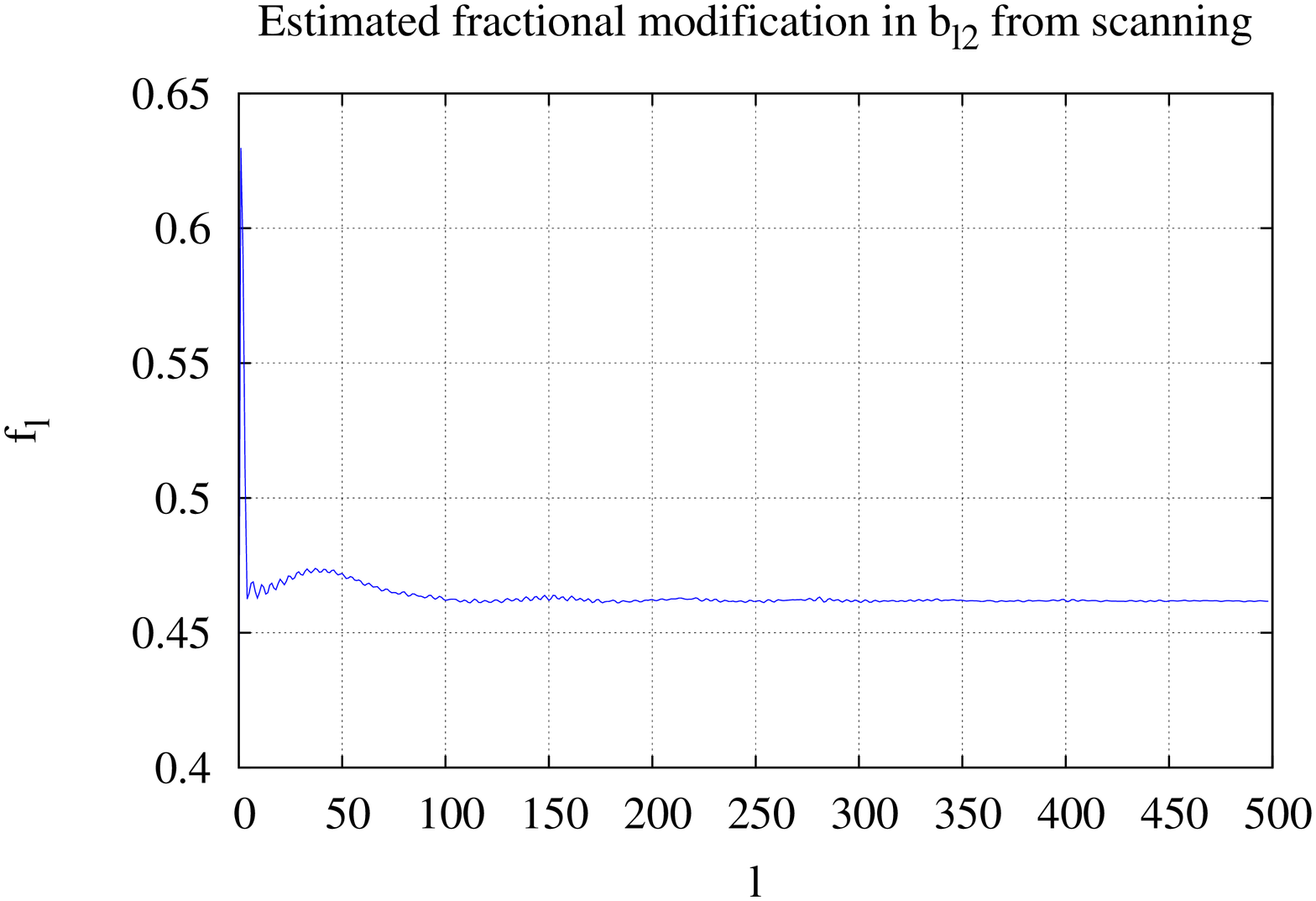}
\caption{Estimation of violation of SI for non-rotating beams can also  provide approximate estimate for more complicated scans in a fairly straightforward way. Using a method described in the text, one can evaluate a multipole dependant scale factor from the scan pattern only (without the beams), which can modify the spherical harmonic coefficient of the beam $b_{l2}$ to provide an effective window function. For WMAP scan the window function is plotted in this figure. As seen in the plot, a constant factor can provide the scale factor for most of the multipoles. When we use this factor, $\sim 0.45$, to predict BipoSH expected from WMAP observations using the semi-analytic results, we do get very good consistency with the measured results.
\label{fig:beamavghits}}
\end{figure}

The analytic expression for $f_l$ in Appendix~\ref{beamavghits}, for
simplicity, pertains only to a single beam scan case. In particular,
the approach cannot account for any complexities that may be
introduced to the effective NC-beam that is used to estimate the BipoSH
spectra arising from linear algebraic map-making procedure using two
beam differencing. However, it motivates the exploration to obtain a
phenomenological effective beam $b^{\rm eff}_{lm} = \sum_{m'} F_{lm'}
b_{lm'}$ expressed in terms of the beam-SH obtained from the raw beam
maps. We proceed with the assumption that the WMAP BipoSH measurements
arise from NC-beam effect and take the 'phenomenological' approach to
determine $b_{l2}$ from BipoSH measurements.  Limiting attention to
the leading order contribution at $m=2$ we parametrize the function
$f_l\equiv F_{l2}$ and fit to the measured BipoSH spectra. In
section~\ref{BipoSHbl2fit}, we first fit for a constant value $\alpha$
that closely reproduces the WMAP BipoSH measurements. While we
find that $\alpha\sim 0.4$ is the best-fit value, the goodness of fit
is not very satisfactory.  We find that a linear correction to the
constant scaling, $b^{\rm eff}_{l2} = (\alpha + \beta\cdot l) b_{l2}$, which
provides very good fits to the WMAP BipoSH measurements.  Numerical
simulations incorporating complexities of two beam differencing and
map-making procedure and deviations from PT-scan approximation are
performed below to decide whether the parameters of the phenomenological
scaling obtained are justified.

\subsection{BipoSH coefficients from single-side WMAP raw beam map}
\label{BipoSHWMAPrawbeam}

\begin{figure*}[!htbp]
\includegraphics[width=0.9\textwidth]{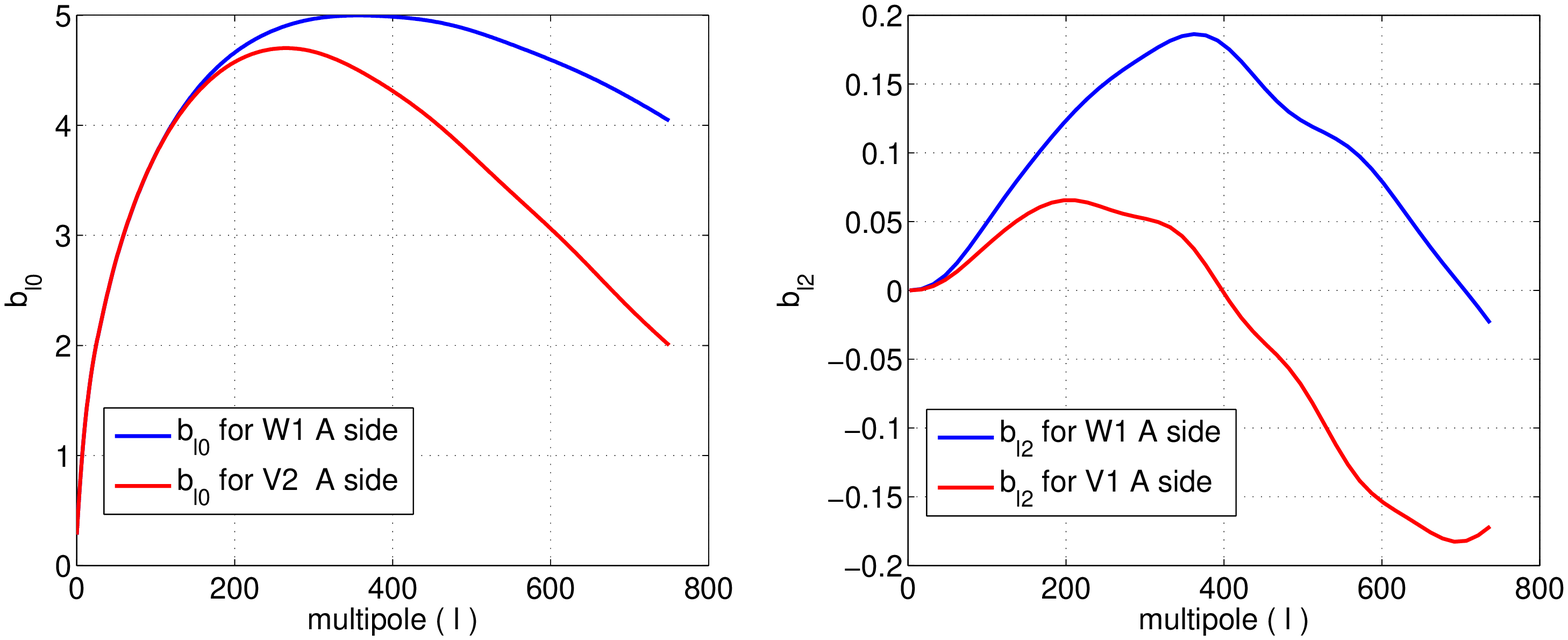}
\includegraphics[width=0.9\textwidth]{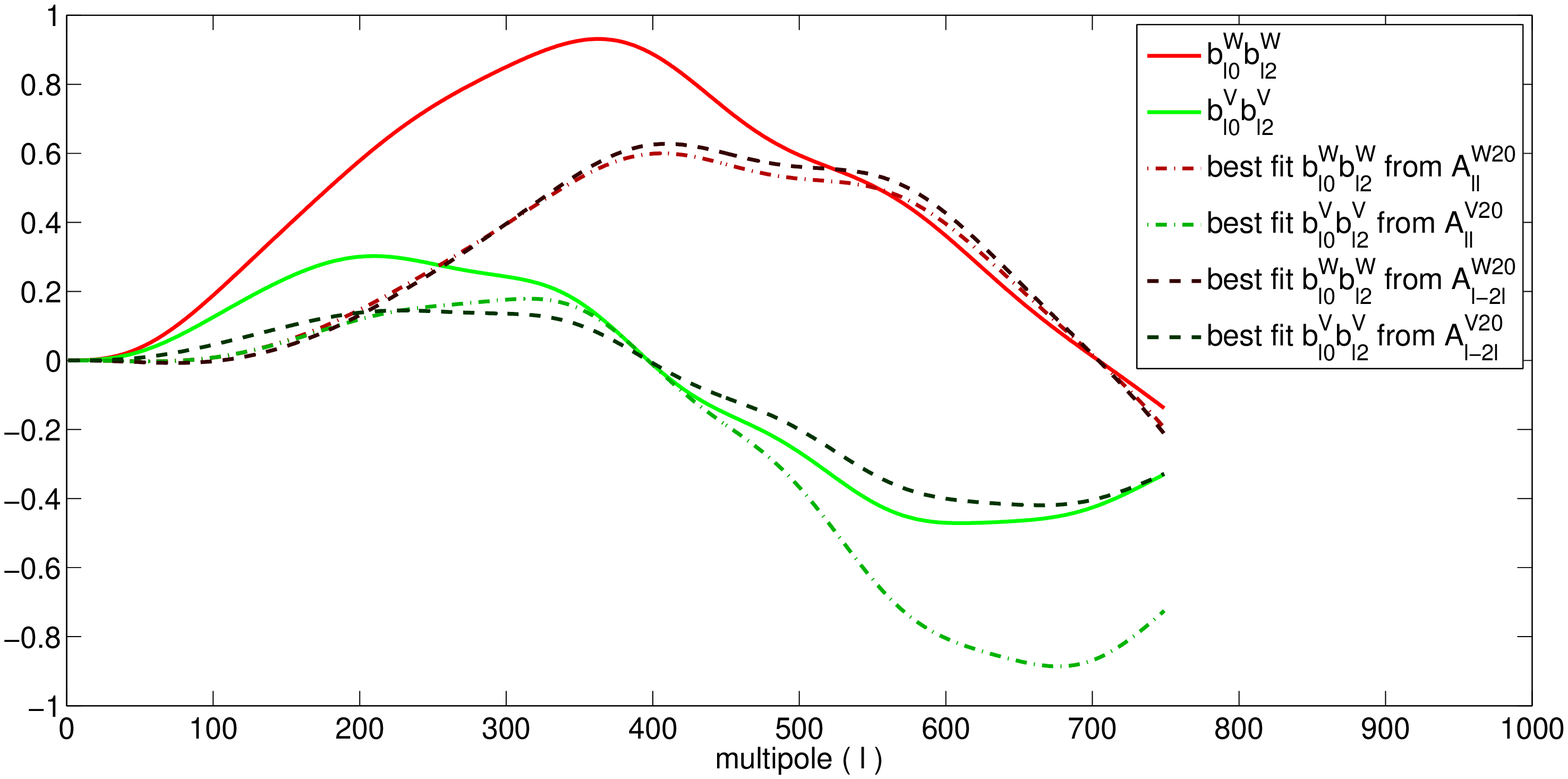}
\caption{ {\em Top:} Beam spherical harmonic transforms, $b_{l0}$ and
$b_{l2}$ of beam maps of A side of W1 and V2 DA. It is interesting to
note that $|b_{l2}|/b_{l0} \lsim 0.01$ implying the BipoSH
representation is sensitive to really really subtle levels of
non-circularity in the beams. \newline {\em Bottom}: The plot of the NC-beam leading
order NC beam perturbation parameter $b_{l0} b_{l2}$ vs. $l$ . The
solid lines correspond to beam-SH of the raw beam maps, the dashed
lines show the $b_{l0} b^{\rm eff}_{l2}$ of the phenomenological
best-fit effective beam-SH $b^{\rm eff}_{l2}=f_l b_{l2}$ obtained by
parametrized linear fits to $f_l$ using the the two measured BipoSH
spectra, $A^{20}_{ll}$ and $A^{20}_{l-2 l}$ The plots also show that
although BipoSH peak structure is largely set by the underlying
angular power spectrum $C_l$ of SI cosmological mode, small
differences observed at the two different frequencies can arise
because of the difference in the shapes of $b_{l0} b_{l2}$. }
\label{fig:blm-W1-V2}
\end{figure*}

\begin{figure*}[!ht]
\includegraphics[width=0.45\textwidth]{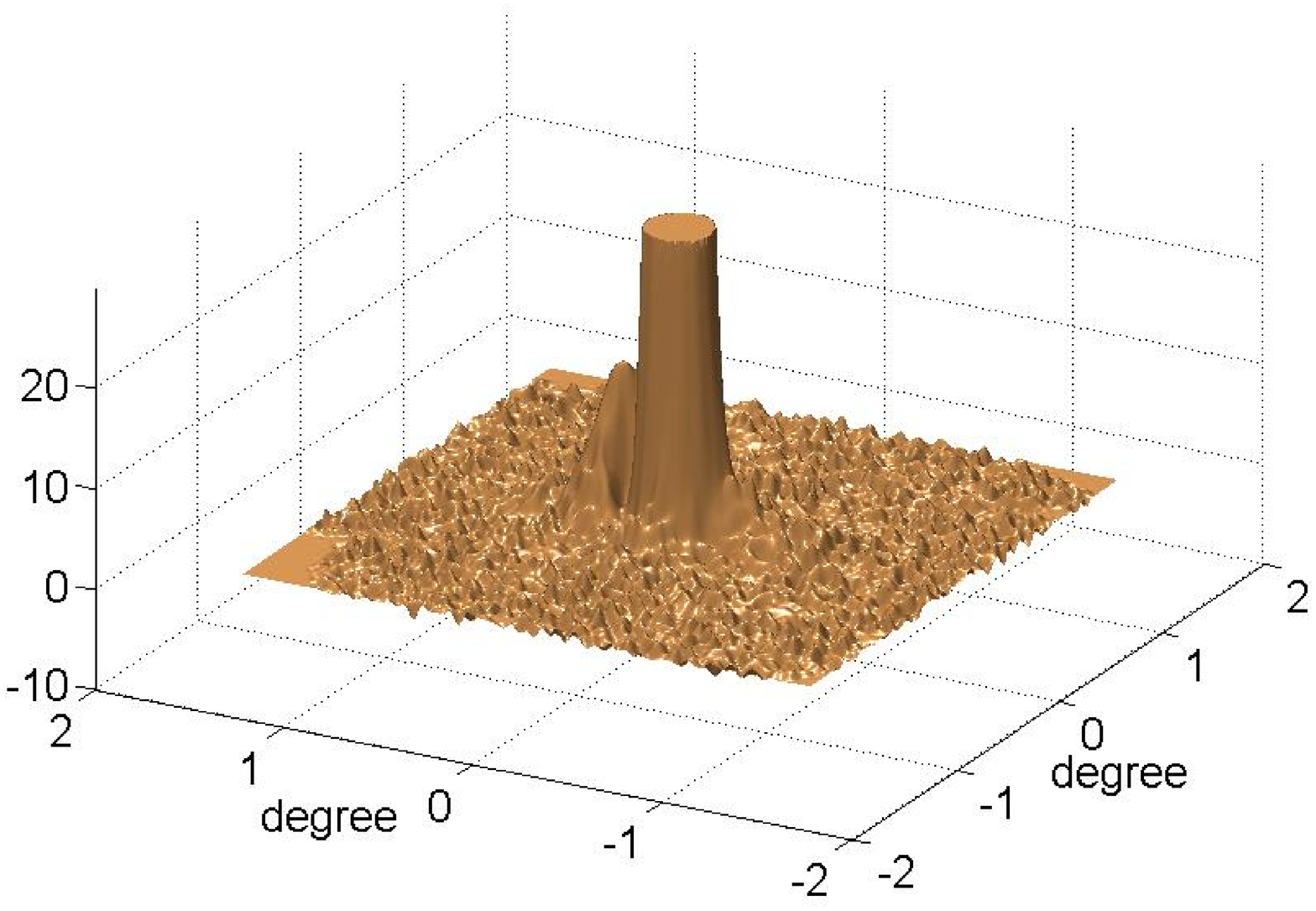}
\includegraphics[width=0.45\textwidth]{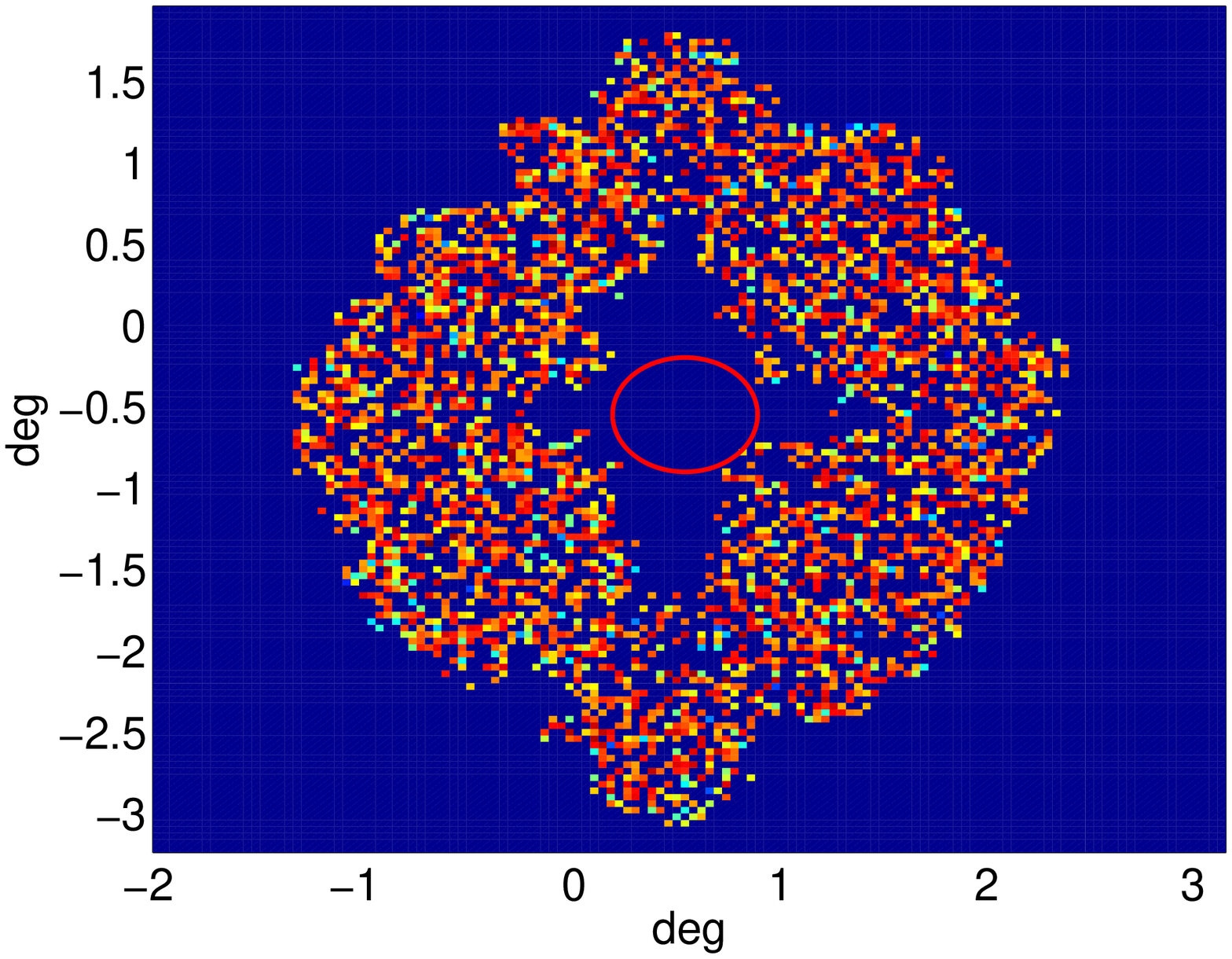}
\includegraphics[width=0.45\textwidth]{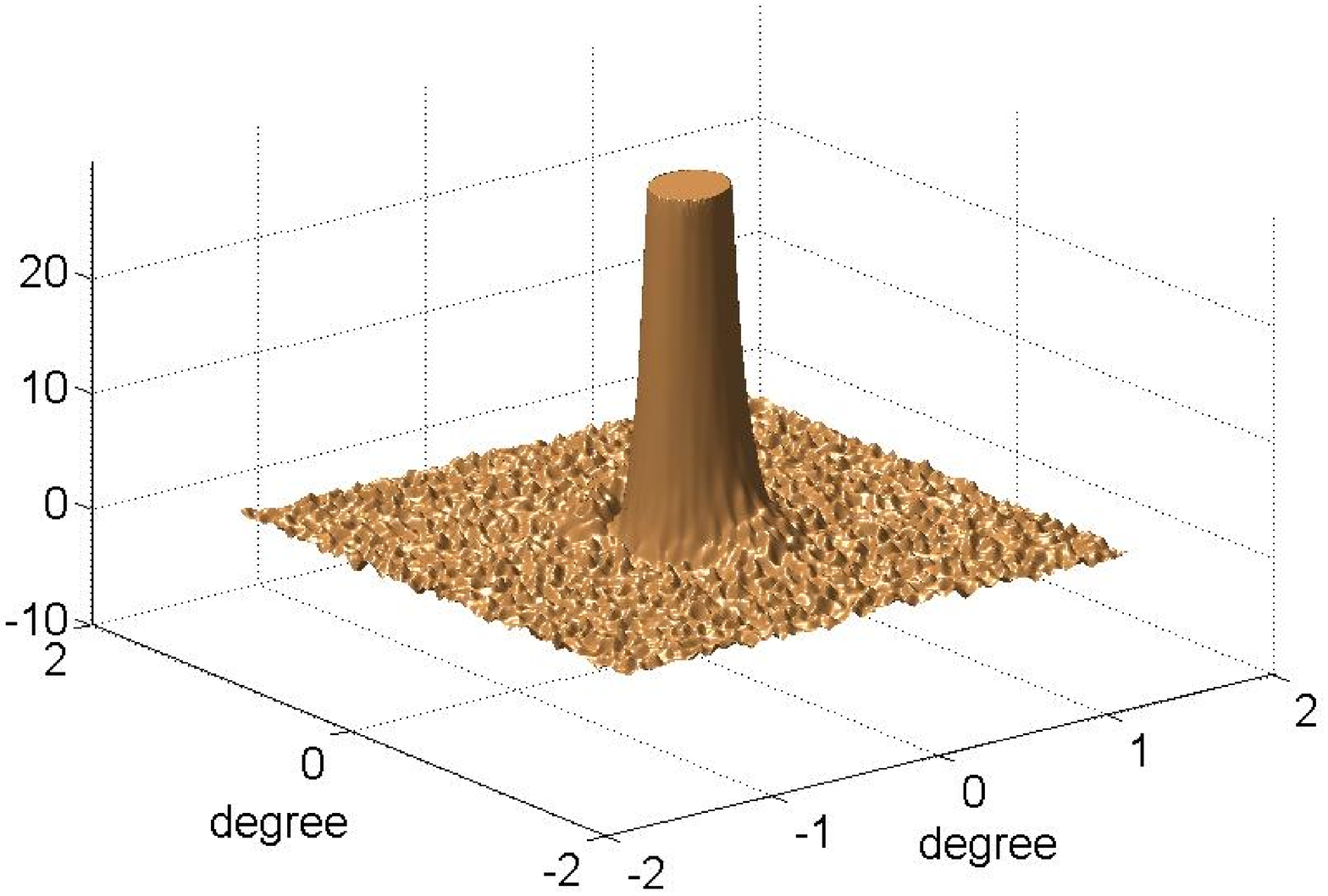}
\includegraphics[width=0.45\textwidth]{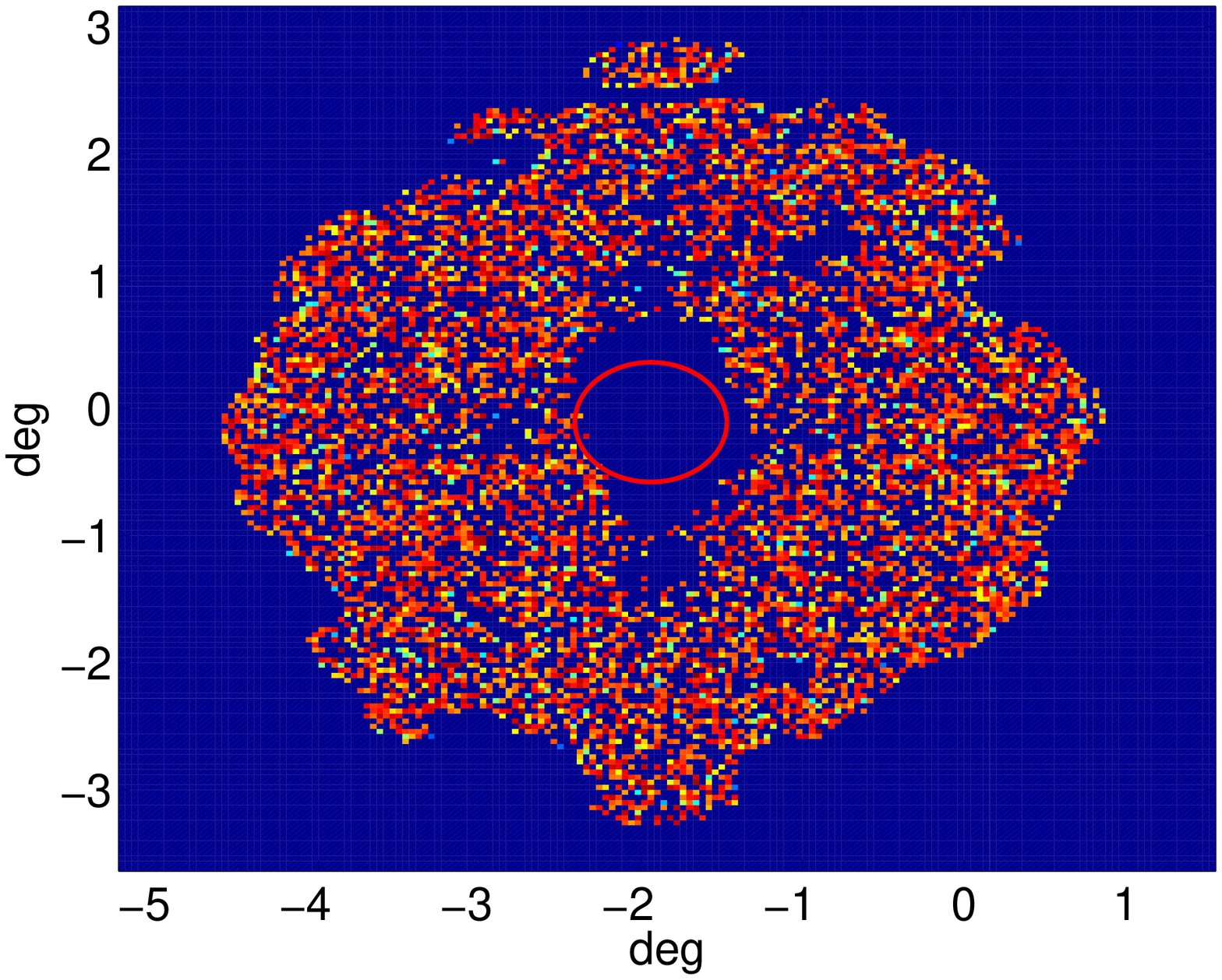}
	 \caption{Beam response function of A side of W1 differencing
assembly {\em (Top)} and A side of V2 differencing assembly {\em
(Bottom)} are shown. The {\em Left-hand} panels show 3D zoomed view of
the central part of the beam shows a truncated view of the central
beam peak to clearly highlight the elliptical contour. Marked
NC shoulder features are seen in the central peak. The {\em
Right-hand} panels cover the entire beam-map images to show the
spread-out annular distribution of regions with negative response.
(The red circles marks the central beam peak region). The power in
negative response is $\sim 0.5$ of the positive power in the central
peaks. The annular negative power distribution shows quadrupolar
feature that modifies the beam-SH $b_{l2}$ to take negative values at
high $l$ for the V-band (see Fig~\ref{fig:W1-V2fullblm}). The
corresponding impact of this beam-SH is significant and crucial for
understanding the intriguing qualitative zero-crossing feature in the
V-band BipoSH spectra.  It is apparent then that correctly accounting
for WMAP NC-beam effects numerically in our ongoing analysis, requires
the convolution of almost the entire beam map region with the SI
sky-map leading to enormous increase in computing costs (relative to
using only the central peak).}
\label{fig:BRA-W1V2}	 
\end{figure*}

To mimic NC-beams closer to WMAP, we consider the A side raw beam maps
of the V2 and W1 differencing Assembly (DA) of WMAP as representative
of the V and W band beams, respectively. We compute the beam-SH
coefficients for these assemblies numerically for use in semi-analytic estimate
of the CMB BipoSH coefficients,using Eqs.~(\ref{eq:WMAP1}) and
(\ref{eq:WMAP1}). Fig.~\ref{fig:blm-W1-V2} is a plot of the
circular, $b_{l0}$ and leading order $b_{l2}$ beam-SH coefficients of
the W1A and V2A raw beam maps. Note that the $b_{l2}$ spectrum changes
sign and takes negative values at large $l$ -- a key qualitative
feature of the WMAP beams that cannot be captured in
Elliptical-Gaussian models where $b_{l2}$ does not change sign with
$l$. The origin of this curious feature becomes apparent in the raw
beam maps of the W1A and V2A channel shown in Fig.~\ref{fig:BRA-W1V2}. The
central part of the beam maps show an elliptical peak with marked
non-trivial NC `shoulder-like' features. However, more
interesting are the right-hand panels, where we highlight a spread out
annular distribution of regions with negative response. The integrated
power in the negative beam response is $\sim0.5$ of the total power and
has significant impact on the beam-SH. In particular, it leads to
negative values of $b_{l2}$ at $l\gsim 2 \sigma_{b}^{-1}$. We find
that this feature is critical in recovering the qualitative feature in
V-band $A^{20}_{ll}$ changing sign at large $l$ seen in the WMAP
measurements. Since such an unique correspondence between a beam-SH
feature and the consequent CMB BipoSH is unlikely to be mimicked by
other effects. We claim that this provides a strong hint that WMAP
BipoSH measurements are linked to the NC-beam effect.

\begin{figure*}[h]
\includegraphics[height=0.45\textwidth, angle=-90]{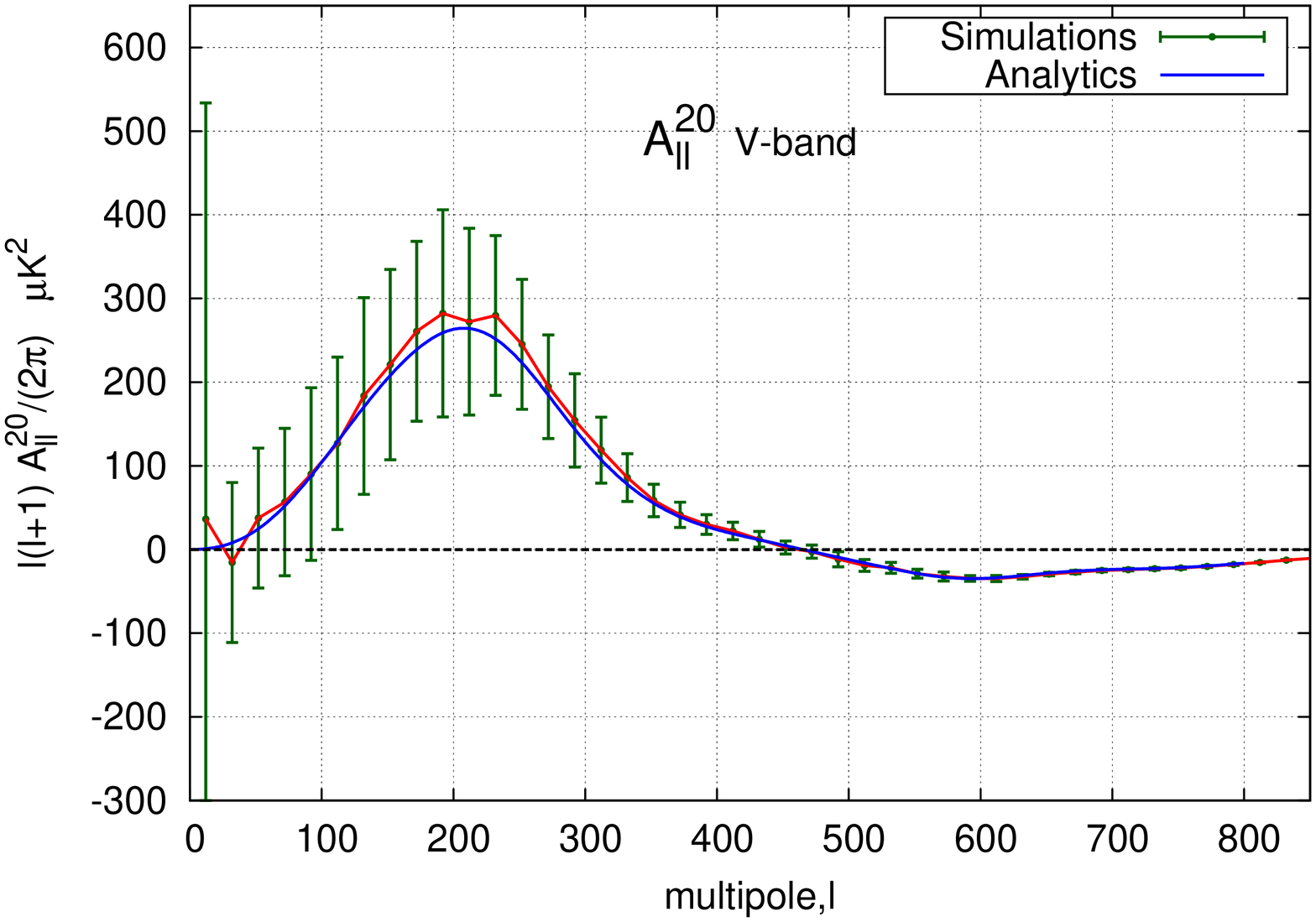}
\includegraphics[height=0.45\textwidth, angle=-90]{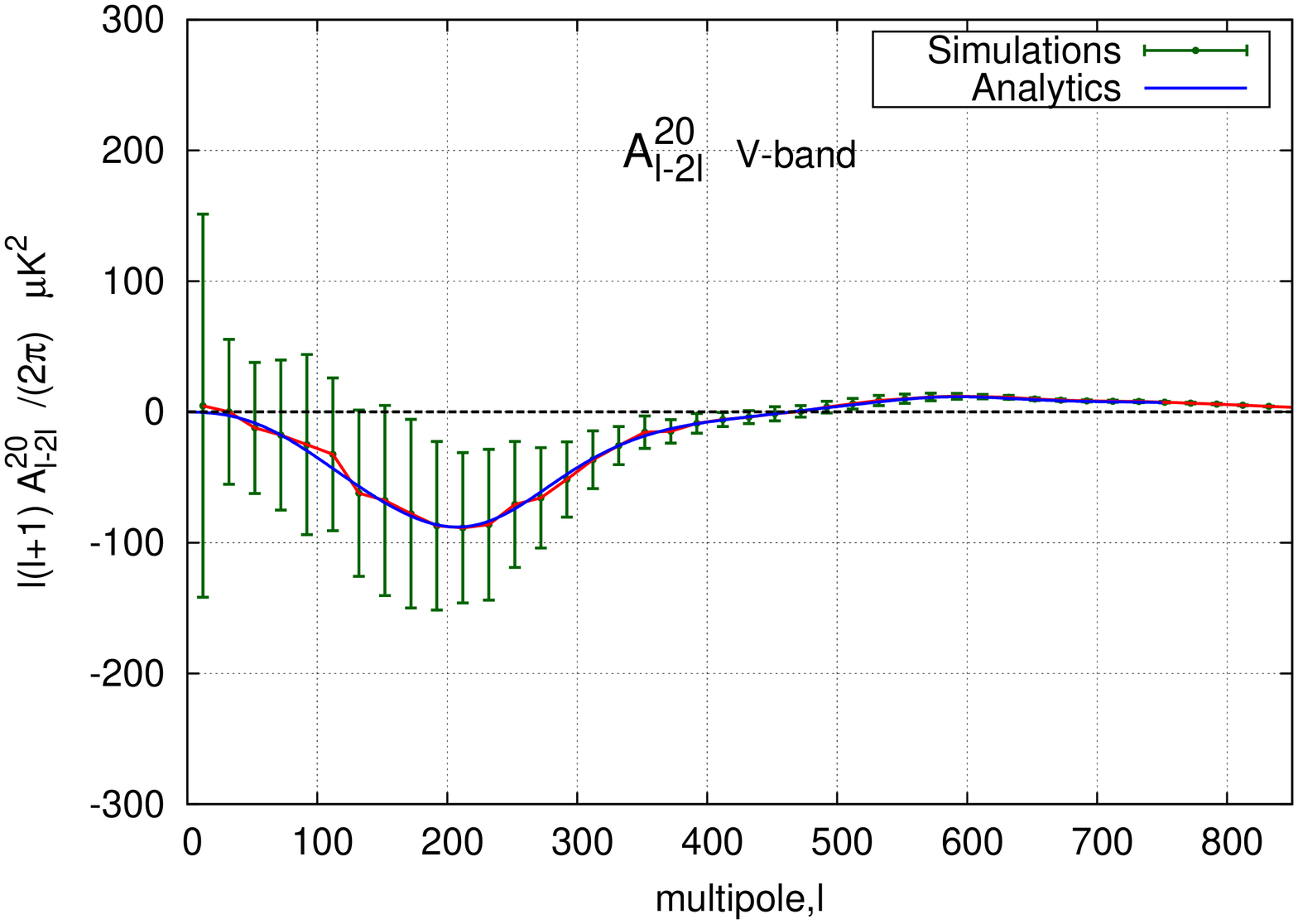}
\includegraphics[height=0.45\textwidth, angle=-90]{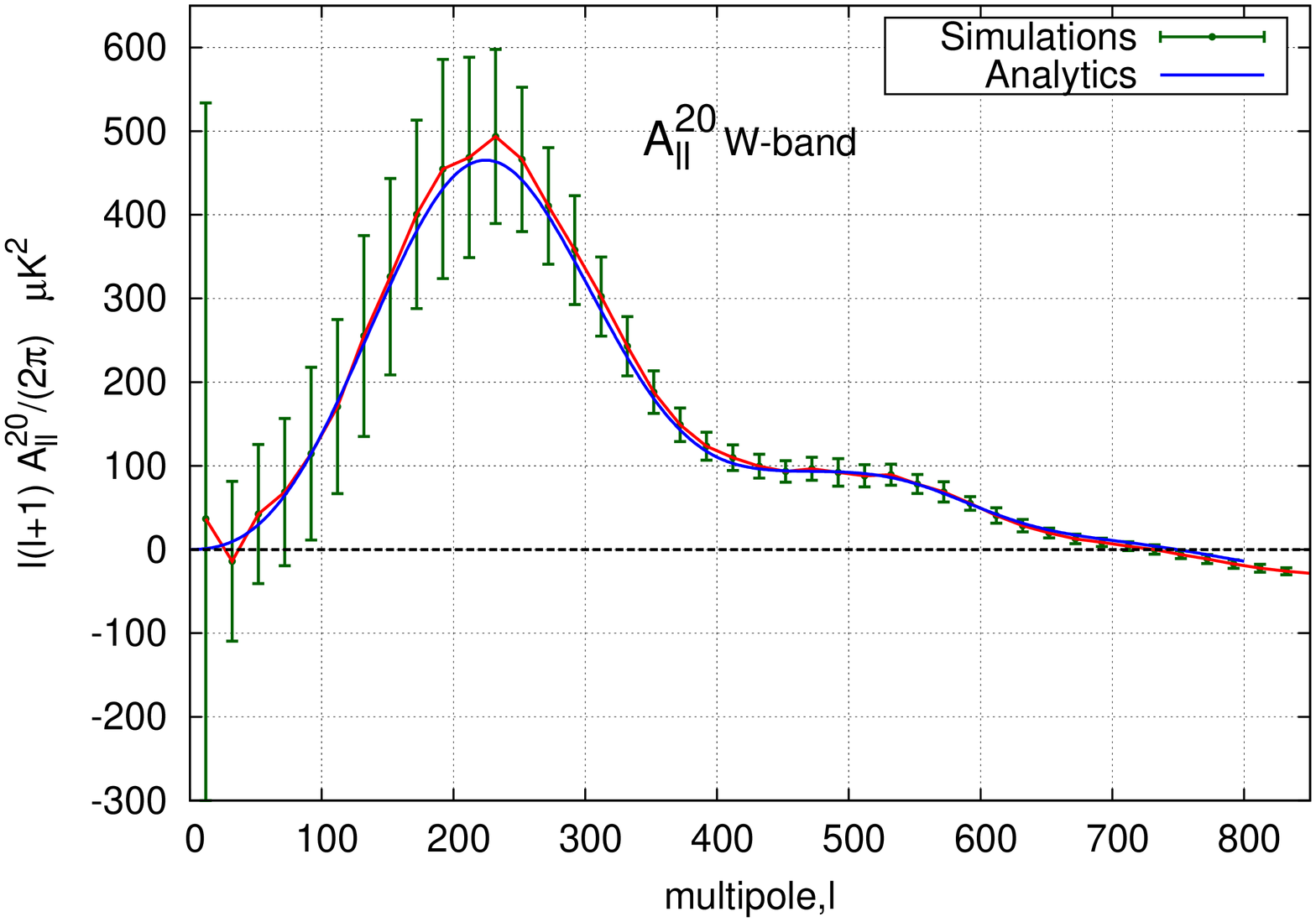}
\includegraphics[height=0.45\textwidth, angle=-90]{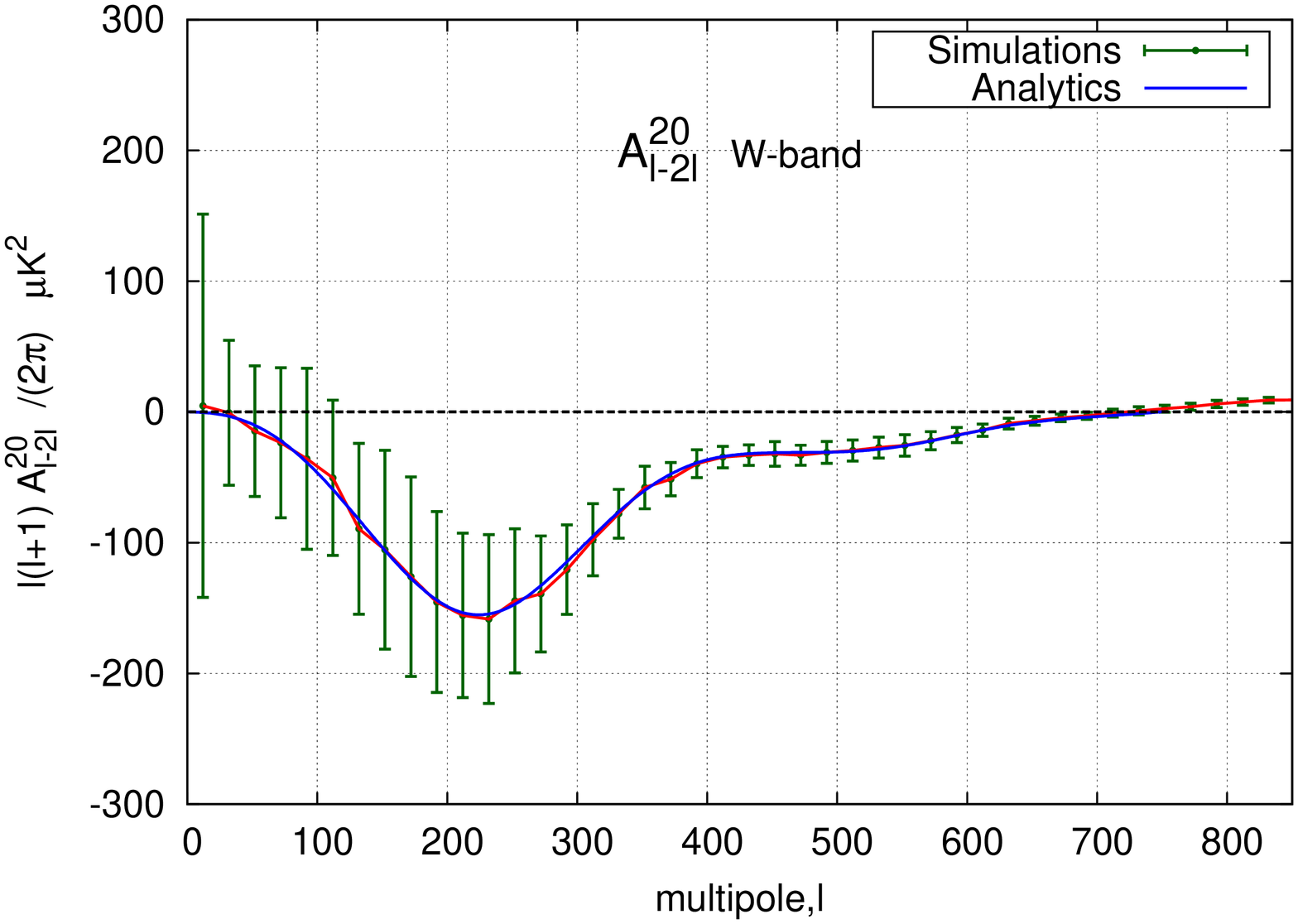}
\caption{BipoSH spectra, $A^{20}_{ll}$ ({\em Left}) and $A^{20}_{l-2
l}$ ({\em Right}) obtained for the raw beam maps A side of V2
channel ({\em Top}) and A side of W1 channel ({\em Bottom}) of the
WMAP experiment.  Analytically evaluated BipoSH spectra (blue)
overlaid on the average BipoSH spectra (red) with error bars obtained
from 100 simulations of statistically isotropic CMB sky convolved W1
and V2 channel of WMAP. While the BipoSH spectra from raw beam-SH
recover interesting qualitative features of the measure spectra, the
amplitudes need to be corrected for the circularizing effect of
averaging raw beam maps with multiple hits with varying orientation at
any pixel.}
\label{fig:W1-V2fullblm}
\end{figure*}

We use PT-scan approximation in ecliptic coordinates given the fact
that the WMAP BipoSH results indicate azimuthal symmetry of SI
violation ($M=0$) in ecliptic coordinates. The analytical results are
verified using numerical simulations of the BipoSH coefficients
$A^{20}_{ll}$ and $A^{20}_{l-2 l}$ are shown in
Fig.~\ref{fig:W1-V2fullblm}.  As seen in Fig. \ref{fig:W1-V2fullblm},
the amplitudes of the BipoSH coefficients do not numerically match WMAP7
year's observed detections. This is expected as the BipoSH
coefficients are obtained from the A side beam of a single channel in
W and V band without accounting for circularizing of the beam due to
multiple hits with varying orientations (this case has been considered later).  We collate and list some key
similarity of these results to the detections of SI violation in
WMAP-7 data:

\begin{enumerate}

\item{} As expected from analytic understanding, the numerical
simulations also confirm that the most significant contribution is the
$M=0$ mode in BipoSH spectra $A^{2M}_{ll}$ and $A^{2M}_{l-2 l}$ when
the analysis is done in ecliptic coordinates, giving strength to our
PT-scan approximation. In a coordinate system where PT-scan
approximation is valid only $M=0$ mode should be significant.
\item{} We notice the NC beam effect is larger in W band than in V
band explaining the difference in detected SI violation signal at the
two frequencies.
\item{} We recover the change of sign of the BipoSH $A^{20}_{ll}$
measurements at large $l$ in the V-band.
\item{} The BipoSH coefficients from NC beam shows a prominent bump
roughly around the first acoustic peak ($l=220$) for both W and V
band. This corresponds mainly to the scale picked by the underlying
angular power spectrum $C_l$. However, the precise peak location also
depends on the peak in $b_{l0} b_{l2}$ for each band and can account
for differences in the peak location in the two bands shown in
Figure~\ref{fig:blm-W1-V2}.
\end{enumerate}

The exercise of computing BipoSH from single side raw beam maps under
PT-scan approximation does provide motivation for a deeper analysis of
the NC-beam effect in the WMAP-7 maps. Numerically simulated maps
incorporating the two side WMAP beam differencing, scan and map-making
details should settle the extent to which approximations used here are
valid and clearly settle the question of whether the entire WMAP-7
BipoSH signal arises as an effect of uncorrected NC-beam effects.
Implementing numerical simulation with actual WMAP beam maps present
serious computational challenge. Besides the main central peak having
significant structure, the WMAP beams have a spread-out annular pattern
of negative response that cannot be neglected. This implies that for a
careful BipoSH analysis of SI violation, a convolution with much
larger regions of the WMAP beam maps of W and V band, extending
roughly to, $\sim 10\theta_{FWHM}$ is required, making it
computationally expensive. The beam maps of W1A and V2A channel are
shown in Fig.\ref{fig:BRA-W1V2}. In particular, we highlight the
spread-out negative part of the beam response that explains some
unique qualitative features of the WMAP BipoSH measurements.
The power in the negative part of the beam is more than 50 percent of
its central positive peak and hence, non-negligible.

\subsection{Fitting BipoSH spectra measurements to an effective NC-beam}
\label{BipoSHbl2fit}

In the analytic approach we show that the BipoSH coefficients scale
as $A^{20}_{l_1 l_2}\propto b_{l 0}b_{l 2}$ for mild deviations
from circular beams under PT-scan as given in
Eq.~(\ref{biposhbeamM0}). However, realistically, complex details of
relative orientation of A and B side of beams for each DA, the
differencing scheme and map-making, the varying orientations for the
multiple hits at each pixel in the actual scan strategy, etc., all
could have bearing at finer levels on CMB BipoSH generated by the
NC-beam.

As shown in Appendix~\ref{beamavghits}, the averaging of the NC-beam
at any pixel due to multiple-hits at different orientations can be
expected to lead to a simple scaling of the form $b^{\rm eff}_{l2} = f_l
b_{l2}$. This would correspondingly scale BipoSH spectra obtained
from A side beam of W1 and V2 channel by $f_l$.

Retaining PT scan approximation, the superposition of raw beam
transforms leads us to expect a constant scaling $f_l=\alpha$.  The
value $\alpha\sim 0.45$ most closely reproduces the WMAP BipoSH
spectra.  We confirm this to be consistent with numerical estimates
from beam-SH superposition using $\rho_i(\hat n)$ obtained from
WMAP-like multiple hits but with PT-scan approximation (see
Appendix~\ref{beamavghits}. The numerical value of $\alpha$ depends on
the relative orientations of the A and B side beams (assumed here
to be identical) and matches the fit value of $\alpha$ for relative
orientation $\sim 140^\circ$.

However, the goodness of fit for constant $f_l$ is not very
satisfactory in explaining the entire WMAP-7 BipoSH spectra in terms
of NC-beam systematics and suggests that $l$-dependence of $f_l$ may
be important for a finer match. While, in principle, it is possible to
determine $f_l$ for the WMAP scan by numerical superposition of
beam-SH using Eq.~(\ref{eq:beamavghits}). Practically, it will be
plagued by ambiguity of relative orientation of the A and B side
NC-beams and many other finer details.  Hence, we prefer to take a
phenomenological approach and explore the possibility of parametrized $f_l$ that fit the
BipoSH measurements. Minimally, this provides an effective beam
that would explain the observed WMAP-7 BipoSH spectra under PT-scan
approximation. We believe that this provides a computationally
inexpensive approach to quantify the WMAP NC-beam systematic effect.

\begin{figure*}[!h]
\includegraphics[height=0.9\textwidth,angle=-90]{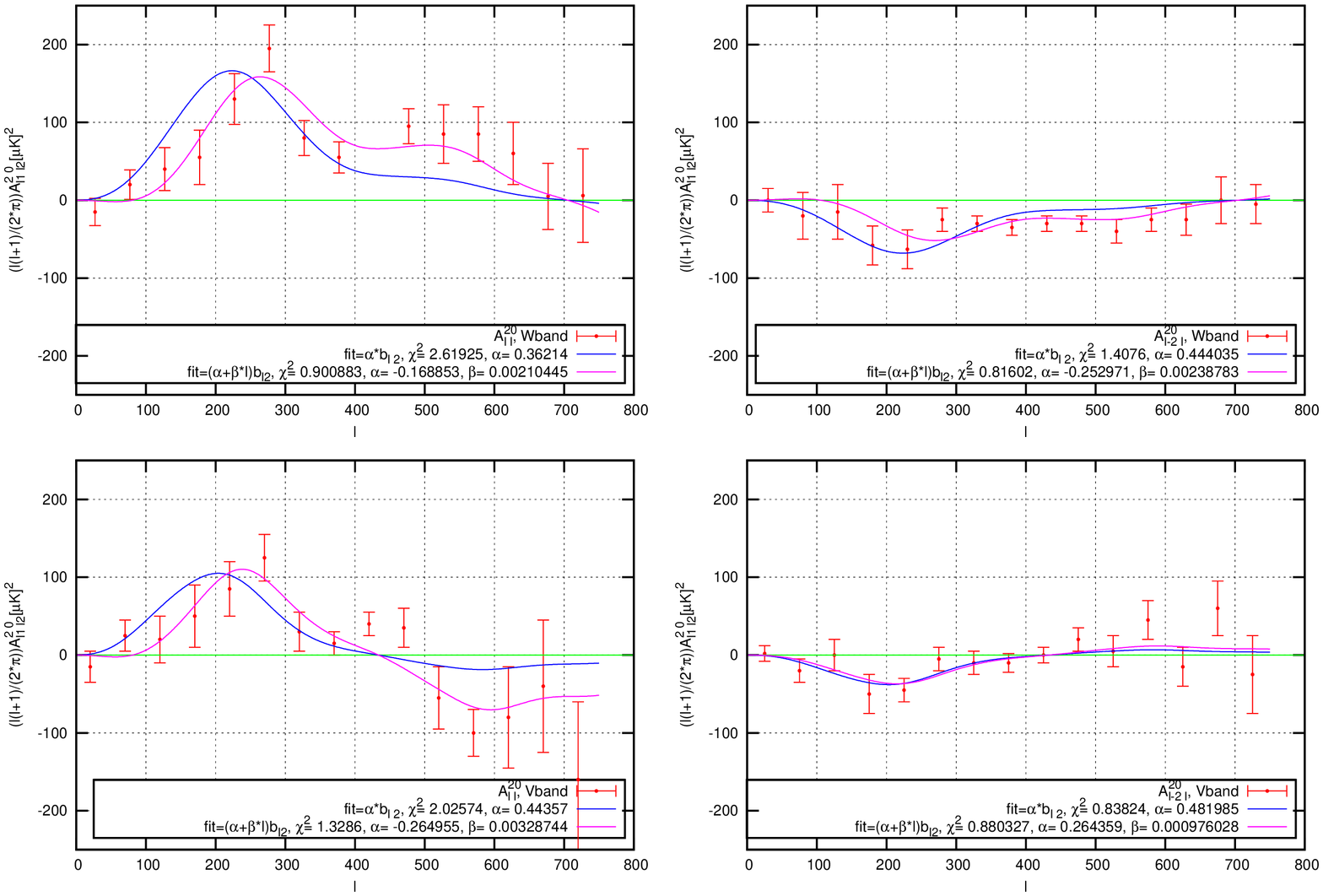}
\caption{Comparison of BipoSH spectra generated by effective NC-beams
 at the V and W bands to the published WMAP-7 BipoSH spectra
 measurements (approximately digitized from the Fig 16 of
 Ref.~\cite{CB-RH-GH}). As described in Appendix~\ref{beamavghits}, an
 effective NC-beam can be constructed from from the raw beam-SH, $
 b_{lm}$ given by $b^{\rm eff}_{lm} = \sum_{m'}F_{lm'} b_{lm'}$.  The
 measured BipoSH spectra, $A^{20}_{ll}$ and $A^{20}_{l-2 l}$, are
 sourced at the leading order by the $m=2$ term. Hence we estimate the
 parametrized fits to the function $f_l \equiv F_{l2}$ using the W and
 V band BipoSH measurements. We present the best fit results for a
 constant $ f_l = \alpha$ and a linear function $f_l =
 (\alpha+\beta\cdot l)$.  The respective values of the reduced
 $\chi^{2}$ of the fit and the best-fit parameters are provided in the
 figure. The constant, $f_l$, provides a fit that partially accounts
 for the BipoSH spectra detections in terms of WMAP NC-beam
 effect. However, remarkably an effective $b^{\rm eff}_{l2}$ from the
 linear form of $f_l$ can satisfactorily explain the non-zero BipoSH
 spectra detected in WMAP-7 in terms of the corresponding V and W band
 effective NC-beams.}
\label{fig:eff-BipoSH}
\end{figure*}

The obvious step beyond constant scaling, would be to add a linear
correction $ f_l = \alpha + \beta l$. As shown in
Fig.~(\ref{fig:eff-BipoSH}) we find very good fits to the WMAP BipoSH
measurements with this scaling.  Interestingly enough, it seems
possible to find values of $\alpha $ and $\beta$ that simultaneously
provide fairly good fits to both $A^{20}_{ll}$ and $A^{20}_{l-2 l}$
BipoSH spectra.  Note that the product $b_{l0} b_{l2}$ still 
tends to zero for large $l$ as seen in the bottom panel of
Fig.~\ref{fig:blm-W1-V2}.

\section{Recovering WMAP7 measured BipoSH with actual scan}

In order to extract BipoSH coefficients in the previous section, we simulated maps
by convolving a statistically isotropic map with the WMAP W-Band and
V-Band beam over the sky. For these simulations, we used a very simplistic
scan pattern, where beams are parallel transported across the sky,
and the angle between the major axis of the elliptical beam and the local
meridian does not vary across the sky. These simulations require
less computation time to provide the scanned maps. BipoSH coefficients obtained from these simulations match reasonably well
with the observed BipoSH of WMAP-7. Evaluating the actual effect coming from
the WMAP scan strategy requires convolving the sky with the real WMAP
beam followed by the actual WMAP scan procedure, map
making procedure and masking as used by WMAP-7.
In this section we have presented the BipoSH coefficients obtained from these simulated maps following the actual WMAP procedure.
Since the entire process is computationally intensive, only 10 simulations
using the W1-Band beam maps are generated and the have evaluated BipoSH coefficients from these many simulations~\footnote{We would be presenting results of more detailed 
simulations in an upcoming publication}. It can be seen that the results match the WMAP-7 non-zero BipoSH observation to a very good accuracy. 

\begin{figure*}
\includegraphics[height=0.45\textwidth,angle=-90]{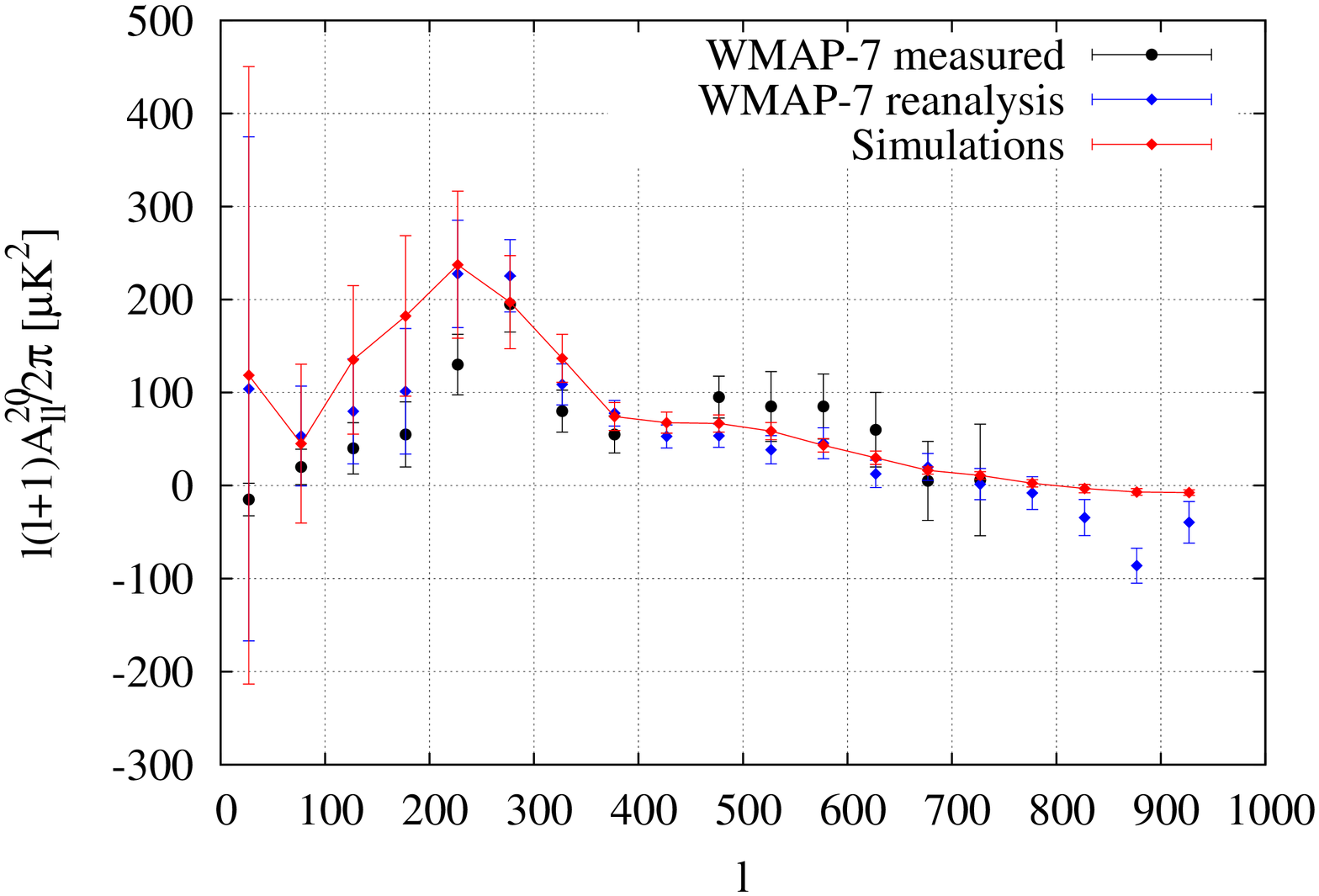}
\includegraphics[height=0.45\textwidth,angle=-90]{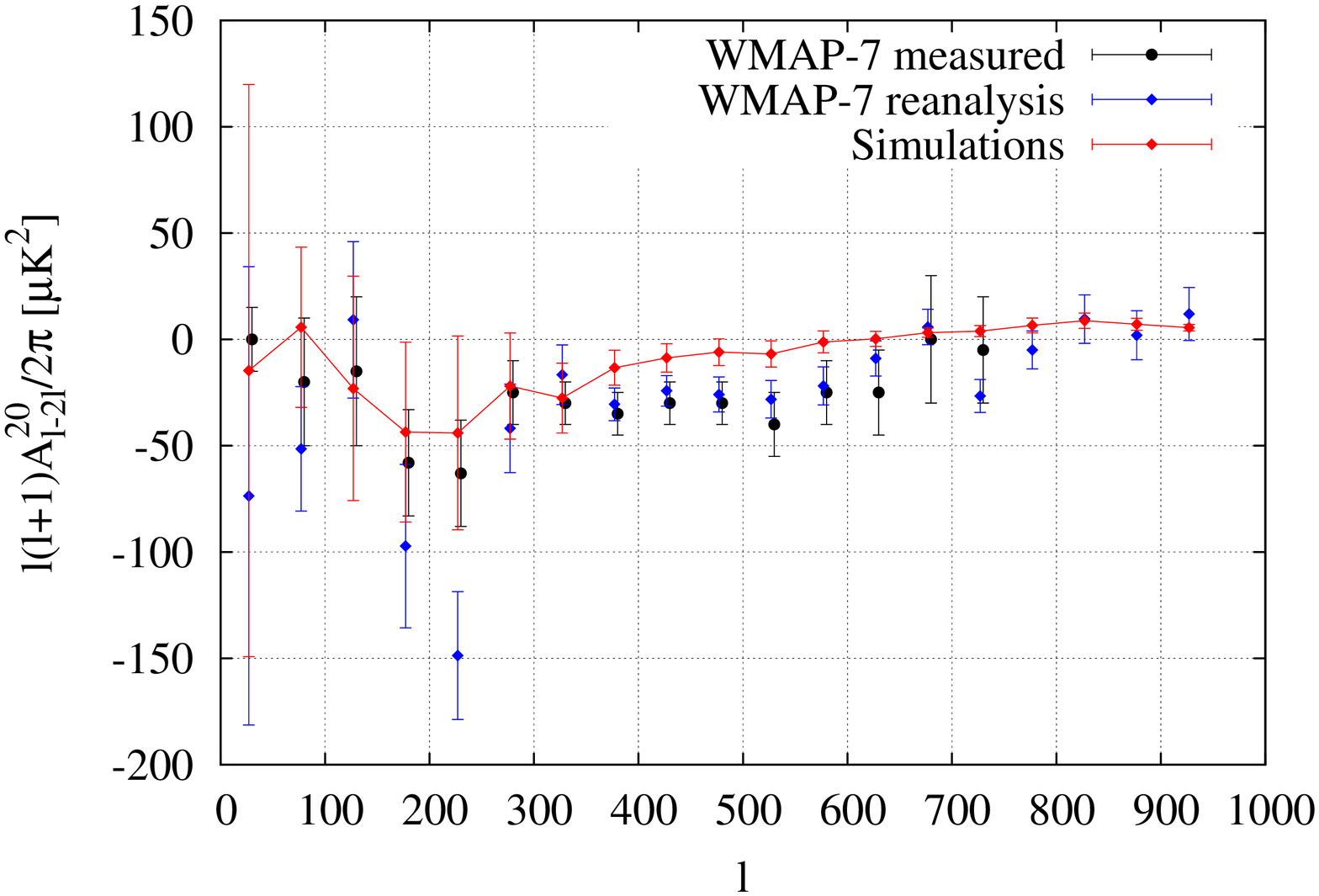}
\caption{To illustrate that the SI violation detected by WMAP can be satisfactorily explained by beam asymmetry, 
we do full simulations of map making for WMAP W1 beam with real scan and real beam. $A^{20}_{\ell\ell}$ (left) and $A^{20}_{\ell-2 \ell}$ (right) of the 
simulated maps (red) are consistent with WMAP measurement (grey) and our reanalysis of WMAP-7 (blue). The difference between the last two occurs due to different 
noise weighting schemes. 
Results with full simulations with all the detectors will appear in a future publication.\label{fig:Full_scan}}
\end{figure*}

Figure~(\ref{fig:Full_scan}) shows a comparison of the computed $A_{\ell\ell}^{20}$ and $A_{\ell-2 \ell}^{20}$, to that measured in WMAP-7 maps. 
The exercise shows that the WMAP-7 observed SI violation
can be completely explained by the effect of NC beams.
It can be seen that at low $\ell$'s the WMAP-7 detections are lesser then our
simulated results. Our re-analysis to compute the BipoSH coefficients for WMAP-7 maps shows that
the power at low $\ell$ is actually more than that reported by the WMAP team.
To check this effect a more detailed analysis
is being carried out. The error bars plotted on the simulated BipoSH
are calculated from the sum of the variances from the full scan simulation
error bars and the error bar from the masking. No noise is added to
the simulations, as we are only interested in the mean effect.
As the error bars are calculated only from 10 realizations, they are
not very accurate. To get a more accurate simulations and to know
the effect of different beams from all four W-Band and two V-Band
maps simulations are ongoing. The full detail of the simulation method
and a thorough analysis will be presented in an upcoming publication.

\section{Discussions \& Conclusion}\label{conclusions}

The observed CMB sky is a convolution of the cosmological signal with
the instrumental beam response function of the experiment (assuming
perfect removal of foreground emissions, etc.). The deconvolution of
the beam effect from the signal is relatively straightforward for an
ideal circular symmetric beam.  Non-Circular (NC) deviations of the
beam, however mild, are practically inevitable in all experiments,
and affect the results obtained at the limits of the sensitivity and
resolution of the recent experiments. The effect of WMAP NC-beam on
the angular power spectrum $C_{l}$ is not very pronounced and is
significant only at very large multipoles. It is very therefore very
interesting that CMB maps obtained with NC-beams disrupt the
rotational invariance of the two point correlation function leading to clearly measurable signatures of 
SI violation.

We show that SI violation measures in the Bipolar Spherical Harmonic
(BipoSH) representation of the observed CMB maps, are well-suited to
capture systematic NC-beam effect. It is important to note that even
though the level of non-circularity in WMAP beams is at $\sim1\%$
percent in $C_{l}$, the BipoSH spectra generated by this effect are measurable.
This points to immense promise and potential of the BipoSH
representation also as a diagnostic tool for current and future CMB
experiments. A key objective is also to assess whether the recent
measurement of non-zero BipoSH spectra, $A^{20}_{ll}$ and $A^{20}_{l-2
l}$ in WMAP-7 in the V \& W band maps~\cite{CB-RH-GH} could arise
from uncorrected NC-beam effects in WMAP-7 maps.

In this paper, we introduce the novel and useful concept of expanding
the NC-beam response function in the BipoSH basis to define
beam-BipoSH coefficients. The beam-BipoSH not only incorporate the
effect of the NC-beam (expressed in terms of the beam-SH -- the
spherical harmonic transform of the beam $b_{lm}(\hat z)$ pointing at
$\hat z$), but also the additional effect of the scan strategy that
determines the beam orientation angle, $\rho(\hat n)$ relative to the
local longitude at any pixel. Neat and completely analytic results for
CMB BipoSH generated by NC-beams can be obtained within the
`parallel-transport' (PT) scan approximation where the beam visits the
pixels at constant $\rho(\hat n)\equiv \rho$. We argue that this
approximation is observationally well-motivated due to the azimuthal
symmetry ($M=0$) in the WMAP-7 BipoSH measurements in the ecliptic
coordinates. All analytical results are verified using corresponding
numerical simulations on SI maps convolved with NC-beam in the
PT-approximation. The numerical simulations also provide estimates of
the error-bars, hence, a measure of the significance of carious
features in the predicted BipoSH spectra.

As a test case we first work with Elliptical-Gaussian (EG) beam with
adjustable eccentricity but FWHM that match the V and W band WMAP
beams.  EG beams have well interpreted NC
parameters and also permits a well-controlled perturbation treatment
for NC deviations ($b_{lm}/b_{l0}$ dies rapidly with
$|m|$~\cite{TS-BR}). This exercise also provides a reliable test
case to cross-check analytic results with that obtained from numerical
simulations of SI maps convolved with NC-beams.  At the leading order
($b_{l2}b_{l0}$), we recover significant CMB BipoSH coefficients
$A^{20}_{ll}$ and $A^{20}_{l-2 l}$ with broadly correct qualitative
features -- relative sign difference in the two spectra and a well
understood peak structure defined by the acoustic features in
underlying $C_l$ modulated by spectral shape of $b_{l0}b_{l2}$. We
choose the eccentricity parameters at fixed FWHM to match the
amplitude. We conclude that the multiple visits of the beam reduces
the eccentricity by a factor comparable to that indicated by comparing
NC beam corrections to WMAP $C_l$ from semi-analytic and
numerical estimates~\cite{SM-AS-TS,Hin_WMAP07,SM-AS-SR-RS-TS}.

As may be expected that the EG approximation to WMAP
beams, completely fails to recover certain qualitative features of the
BipoSH measurements, in particular, the change of sign in the V-band
$A^{20}_{ll}$ at large $l$. This is addressed next by computing the
beam-SH for single (A) side WMAP beam maps for V and W bands. The beam
maps show a marked annular region well beyond a few FWHM from the
center where the beam response is negative. This leads to the $b_{l2}$
going negative at large multipoles beyond the beam-width and explains
the change in sign of V-band BipoSH at larger $l$. Note that
EG beams can never reproduce this feature, since $b_{l2}\ge
0$.  Again, as expected, the amplitude of the CMB-BipoSH coefficients based on
beam-SH, $b_{l2}$, from raw beam maps are much higher. An analytic
treatment of superposition of raw beam transforms to mimic the effect
of multiple hits with varying orientation at a pixel, leads us to
expect a scaling of the leading order term $b^{\rm eff}_{l2} = f_l
b_{l2}$. However, $f_l$ depends on the detailed and accurate scan
description of each beam and many other finer details of the
instruments. Hence we choose to adopt a `phenomenological' approach to
determine the effective beam as a result of multiple hits with different orientations at a pixel 
and fit a parametrized $f_l$ to the
WMAP-7 BipoSH spectra measurements assuming it to be entirely sourced
by NC-beam effect. A constant $f_l=\alpha\approx 0.4$ provides only a
fair fit and does not account for the entire signal. A linear
correction to the constant scaling, $f_l= \alpha + l\dot\beta$ allows
for very good fits to the WMAP BipoSH spectra measurements as shown in
Fig.~\ref{fig:eff-BipoSH}.

Complexities of two beam differencing and map-making procedure,
deviations from PT-scan approximation, etc., may account for the
phenomenologically obtained $f_l$ scaling of raw beam-SH, $b_{l2}$. 
We simulated 10 maps for W1-band following the actual WMAP procedure and obtained BipoSH spectra from them which match well with the observed BipoSH spectra 
in WMAP-7 year data.
We can firmly conclude that
NC-beam effect is dominantly responsible for the WMAP-7 non-zero BipoSH
measurements.  In any case, our results indicate clearly that a major
portion of the BipoSH signal measured in WMAP-7 may be attributed to
an experiment specific systematic effect arising from uncorrected
NC-Beam effect.  However, it is still possible that the BipoSH
measurements are not entirely explained by the NC beam effect leaving
room for an achromatic component from intrinsic cosmological SI
violation.  Since the WMAP-7 BipoSH signal are strong and significant,
other full sky CMB experiments with comparable, or better, sensitivity
and angular resolution should readily confirm, or disprove this
possibility. 
We will be providing full extensive simulation for W-band and V-band and details of effect of non-circular beams on SI violation in future publication.
In near future, good quality full sky CMB polarization
maps when available can also be studied in the BipoSH
representation~\cite{SB-AH-TS} and will provide an independent window
in to SI violation phenomena.

\section*{Acknowledgments}

We simulate CMB maps with the HEALPix \cite{hpix} package additional
modules added for real-space convolution with NC-beams. Computations
were carried out at the HPC facilities at IUCAA.  NJ, SD and AR
acknowledge the Council of Scientific and Industrial Research (CSIR),
India for financial support through Senior Research fellowships. NJ
thanks IUCAA for hosting visits to facilitate the collaborative
effort. NJ also acknowledges useful discussions and encouragement from
Sanjay Jhingan at her home institution. SM acknowledges support from
his DST fast-track grant.  TS acknowledges support from the
Swarnajayanti fellowship grant, DST, India.

\appendix

\section{Beam BipoSH}\label{app:beam-biposh}

Beam-BipoSH are expansion coefficients of the beam response function
in BipoSH basis (see Sec~\ref{ncbeambposh}).  The most general
beam-BipoSH in any coordinate system is given by, 

\begin{eqnarray}
&&B^{LM}_{l_1 l_2} \ =\ -\sum_{m_1 m_2}C^{LM}_{l_1 m_1 l_2 m_2}\sum_{m'}b_{l_2
m'}(\hat z) \ \times\\
&&\quad \int^{\pi}_{0} d(\cos\theta) \int^{2\pi}_0 d\phi D^{l_2}_{m_2
m'}(\phi,\theta,\rho(\theta,\phi))Y^{*}_{l_1
m_1}(\theta,\phi). \nonumber
\end{eqnarray}
 Wigner-\textit{D} functions can be expressed in terms of Wigner-\textit{d} through following relation,
\begin{eqnarray}
D^{l}_{m m'}(\phi,\theta,\rho)={\rm e}^{-i m \phi}\, d^{l}_{m
m'}(\theta)\,{\rm e}^{-i m' \rho}\,,
\end{eqnarray}
and reduces to spherical harmonics for $m'=0$,
\begin{eqnarray}
D^{l}_{m0}(\phi,\theta,\rho)=\sqrt{4\pi/(2l+1)}Y^{*}_{lm}(\theta,\phi)\,.
\end{eqnarray}

In the parallel-transport (PT) scan approximation, the beam
orientation, with respect to the local Cartesian coordinate aligned
with the spherical $(\theta, \phi)$ coordinates, does not vary on sky
(i.e., $\rho(\theta,\phi)\equiv\rho$), the beam-BipoSH coefficients for such an approximation are,
\begin{eqnarray}
B^{LM}_{l_1 l_2}&=&-\sqrt{\frac{2l_{1} +1}{4\pi}}\sum_{m'}b_{l_2
m'} e^{-i m' \rho}\times\nonumber \\
&& \Big(\sum_{m_1 m_2}C^{LM}_{l_1 m_1 l_2
m_2} \int^{2\pi}_0 d\phi e^{-i(m_{1}+m_{2})\phi} 
\nonumber\\ &&\int^{\pi}_{\theta=0}d^{l_2}_{m_{2} m'}(\theta)d^{l_1}_{m_{1} 0}(\theta)d(\cos\theta)\Big).
\end{eqnarray}
Which simplifies to,
\begin{eqnarray}
B^{LM}_{l_1 l_2}&=&
2\pi\delta_{M0}\sum_{m'}b_{l_2 m'}(\hat z) e^{-i m' \rho} \ \times \nonumber\\ && \quad \sum_{m_2}(-1)^{m_2}C^{L0}_{l_1 -m_2 l_2 m_2} I^{l_1
l_2}_{m_2,m'},
\end{eqnarray}
and is derived to be non-zero only for $M=0$, using the orthogonality
relation,
\begin{eqnarray}
\int^{2\pi}_0 d\phi\exp^{-i (m_1+m_2) \phi} \ = \ 2\pi\delta_{m_1,-m_2}.
\end{eqnarray}
We define the notation
\begin{eqnarray}
 I^{l_1 l_2}_{m_2,m'}=(-1)^{m_2 +1} \sqrt{\frac{(2l_{1}+1)}{4\pi}}\times\nonumber \\ \int^{\pi}_{\theta=0}d^{l_2}_{m_{2} m'}(\theta)d^{l_1}_{m_{2} 0}(\theta)d(\cos\theta).
\end{eqnarray}
Here we have used $d^{l}_{m m'}(\theta)=(-1)^{m-m'}d^{l}_{-m -m'}(\theta)$.
To simplify the analytic expressions, we retain only the leading order
NC beam spherical harmonic mode $m'=2$, assuming mild NC-beam with
discrete even-fold azimuthal symmetry where no odd $m'$ modes will
contribute. Hence, the summation over $m'$ has three terms
corresponding to $m'=0,\pm2$.

The beam-BipoSH can be then be written as
\begin{eqnarray}\label{eq:thterm}
B^{LM}_{l_1 l_2} &\equiv & B^{LM^{(C)}}_{l_1 l_2}+B^{LM^{(NC)}}_{l_1 l_2} \\
B^{LM^{(C)}}_{l_1 l_2} &=& 2\pi\,\delta_{L0}\,\delta_{M0} \ b_{l_2 0}(\hat z)\sum_{m_2}C^{L0}_{l_1 -m_2 l_2 m_2} I^{l_1
l_2}_{m_2,0} \nonumber\\
B^{LM^{(NC)}}_{l_1 l_2} &=& 2\pi\delta_{M0} \sum_{m_{2}\neq0}C^{L0}_{l_1 -m_2 l_2 m_2} \ \times \\
\Big( b_{l_2 -2}(\hat z)&\exp^{i2\rho}& I^{l_1 l_2}_{m_2,-2}+ b_{l_2 2}(\hat
z)~\exp^{-i2\rho}~ I^{l_1 l_2}_{m_2,2} \Big). \nonumber
\end{eqnarray}
First term in Eq.~(\ref{eq:thterm}) is the trivial beam-BipoSH,
$B^{00^{(C)}}_{l l}$, corresponding to the circular symmetric component of
the beam response function.  NC part of the beam function
$m'=\pm2$, gives rise to beam BipoSH having $L\neq0$. First, we
evaluate the beam-BipoSH due to circular part of beam function.
Orthogonality of Wigner-\textit{d} functions,
\begin{eqnarray}
-\int^{\pi}_{0} d(\cos\theta) \,d^{l}_{m m'}(\theta)d^{l'}_{m m'}(\theta)=\frac{2}{2l+1}\delta_{l l'}
\end{eqnarray}
implies 
\begin{eqnarray}
I^{l_1 l_2}_{m_2,0}=(-1)^{m_2}\Big(\frac{2}{2l_2+1}\Big)\sqrt{\frac{(2l_{1}+1)}{4\pi}}\delta_{l_1 l_2}. 
\end{eqnarray}

Using the following property of Clebsch-Gordan ,
\begin{eqnarray}
\sum_{m}(-1)^{l-m}C^{L0}_{l m l -m}=\sqrt{(2l+1)}\delta_{L0}.
\end{eqnarray}
we obtain
\begin{eqnarray}
B^{LM^{}(C)}_{l_1 l_2}=\sqrt{4\pi}(-1)^{l_2}b_{l_{2} 0}(\hat z)\delta_{l_1 l_2}\delta_{L0}\delta_{M0}.
\end{eqnarray}
Since,
\begin{eqnarray}
b_{l0}(\hat z)= \sqrt{\frac{(2l+1)}{4\pi}}B_{l},
\end{eqnarray}
where $B_{l}$ is the usual beam transfer function of the
circular-symmetrized beam profile,
\begin{eqnarray}
B^{LM^{}(C)}_{l_1 l_2}=(-1)^{l_2}\sqrt{2l_{2}+1}B_{l_2}\delta_{l_1 l_2}\delta_{L0}\delta_{M0}.
\end{eqnarray}

Next, we analytically calculate the beam-BipoSH $B^{LM^{(NC)}}$ due to
the NC part of the beam function,
\begin{eqnarray}
B^{LM^{(NC)}}_{l_1 l_2} &=&-2\pi\sqrt{\frac{(2l_{1}+1)}{4\pi}}\delta_{M0}\sum_{m_2}(-1)^{m_2}C^{L0}_{l_1 -m_2 l_2 m_2}\nonumber\\
&\times&\Big(b_{l_2 -2}(\hat z)\exp^{i2\rho}
\int^{\pi}_{0}d^{l_2}_{m_{2} -2}(\theta)d^{l_1}_{m_{2} 0}(\theta)
\nonumber\\ d(\cos\theta)
&+& \ b_{l_2 2}(\hat z)\exp^{-i2\rho}\int^{\pi}_{0}d^{l_2}_{m_{2} 2}(\theta)d^{l_1}_{m_{2} 0}(\theta)d(\cos\theta)
\Big)\,.\nonumber
\label{eq:NC-BipoSH}
\end{eqnarray}
In the above expression, the summation is over $m_2$. It is convenient
to separate the calculation of the $m_2=0$ and rest of the
$m_{2}\neq0$ terms.

Consider the integral for $m_2=0$.  For this we use, $d^{l}_{m
m'}=(-1)^{m+m'}d^{l}_{m' m}$ and expansion of Wigner-\textit{d}'s in
terms of associated Legendre polynomials, $d^{l}_{m
0}(\theta)=(-1)^{m}\sqrt{(l-m)!/(l+m)!}P^{m}_{l}(\cos\theta)$, and
obtain
\begin{eqnarray}
&&\int^{\pi}_{\theta=0}d^{l_2}_{0 2}(\theta)d^{l_1}_{0
0}(\theta)d(\cos\theta) \ = \\
&& \quad \sqrt{\frac{(l_{2}-2)!}{(l_{2}+2)!}}
\int^{\pi}_{\theta=0}P^{2}_{l_2}(\cos\theta)P_{l_1}(\cos\theta)d(\cos\theta)\,. \nonumber
\end{eqnarray}
Using standard recurrence relations of Associated Legendre functions,
\begin{eqnarray}
P^{2}_{l}(\cos\theta)=\frac{2\cos\theta}{\sin\theta}P^{1}_{l}(\cos\theta)-l(l+1)P_{l}(\cos\theta),
\end{eqnarray}
\begin{eqnarray}
P^{1}_{l}(\cos\theta)=\sin\theta P^{\prime}_{l}(\cos\theta) ,
\end{eqnarray}
and orthogonality relations,
\begin{eqnarray}
-\int^{\pi}_{0}
P_{l_2}(\cos\theta)P_{l_1}(\cos\theta)d(\cos\theta) \ =\ \frac{2~\delta_{l_1
l_2}}{2l_{2}+1}.
\end{eqnarray}
\begin{eqnarray}
-\int^{\pi}_{0}\cos\theta P^{{\prime}}_{l_2}(\cos\theta)P^{0}_{l_1}(\cos\theta)d(\cos\theta)  \nonumber\\
= \ \left\{
\begin{array}{ll}
 0 & \mbox{if ($l_1 +l_2=\textrm{odd}$)} \\ \\
0 & \mbox{if ($l_1 >l_2$)} \\ \\
0 & \mbox{if ($l_1 <l_2$)} \\ \\
\frac{2l_{2}}{2l_{2}+1} & \mbox{if ($l_1 = l_2$)}  \,.\\
\end{array}
\right.
\end{eqnarray}
Therefore, for $m_2=0$, the integral simplifies to
\begin{eqnarray}
&&I^{l_1 l_2}_{0,\pm2}\ \ = \ (-1)^{m_2}\sqrt{\frac{(2l_1 +1)}{4\pi}}\ \times \nonumber\\
&& \quad\left\{
\begin{array}{ll}
0 & \mbox{if ($l_1 +l_2=\textrm{odd}$)} \\ \\
0 & \mbox{if ($l_1 >l_2$)} \\ \\
4\sqrt{\frac{(l_{2}-2)!}{(l_{2}+2)!}} & \mbox{if ($l_1 <l_2$)} \\ \\
\sqrt{\frac{(l_{2}-2)!}{(l_{2}+2)!}}\big[\frac{4l_2}{(2l_2 +1)}-\frac{2l_2(l_2 +1)}{(2l_2 +1)}\big] & \mbox{if ($l_1 = l_2$)} \,.\\ 
\end{array}
\right.
\end{eqnarray}
\newline Next, we consider the $m_2\neq0$ terms in the summation in
Eq.~(\ref{eq:NC-BipoSH}).  Here, $d^{l_2}_{m_{2} 2}(\theta)$ can be
recursively reduced to $d^{l_2}_{m_{2} 0}(\theta)$ using the following
recurrence relation,
\begin{eqnarray}
d^{l_2}_{m_{2} 2}(\theta) &=& \frac{\kappa}{\sin^{2}\theta}\left[\kappa_{0}d^{l_2}_{m_2 0}(\theta) \ + \ \kappa_{1}d^{l_{2}+1}_{m_2 0}(\theta) \ + \right.\\
&& \quad \left. \kappa_{-1}d^{l_{2}-1}_{m_2 0}(\theta)+\kappa_{2}d^{l_{2}+2}_{m_2 0}(\theta)+\kappa_{-2}d^{l_{2}-2}_{m_2 0}(\theta)\right] \, .\nonumber
\end{eqnarray}
where,
\begin{eqnarray*} 
\kappa_0 \ &\equiv& \ \frac{m^2_{2}}{l^2_{2}(l_{2}+1)^2} \ -\
\frac{l_{2}^2-m^2_{2}}{l_{2}^2(4l^2_{2}-1)} \nonumber\\
&&\quad - \ \frac{(l_{2}+1)^2-m^2_{2}}{(l_{2}+1)^2(2l_{2}+1)(2l_{2}+3)}, \\ \kappa_1 \ &\equiv& \
2m_{2}\frac{\sqrt{(l_{2}+1)^2-m^2_{2}}}{l_{2}(l_{2}+1)(l_{2}+2)(2l_{2}+1)}, \\
\kappa_{-1} &\equiv&  -2m_{2}\frac{\sqrt{l^2_{2}-m^2_{2}}}{l_{2}(l^2_{2}-1)(2l_{2}+1)},\\ \kappa_2 \
&\equiv& \
\frac{\sqrt{[(l_{2}+1)^2-m^2_{2}][(l_{2}+2)^2-m^2_{2}]}}{(l_{2}+1)(l_{2}+2)(2l_{2}+1)(2l_{2}+3)} \\
\kappa_{-2} &\equiv &
\frac{\sqrt{(l_{2}^2-m^2_{2})[(l_{2}-1)^2-m^2_{2}]}}{l_{2}(l_{2}-1)(4l^2_{2}-1)} \,.
\end{eqnarray*}
Under reflection Wigner-\textit{d}'s transform as,
$d^{l}_{m m'}(\pi-\theta)=(-1)^{l+m'}d^{l}_{m -m'}(\theta)$.
\begin{eqnarray}
d^{l_2}_{m_{2} -2}(\theta) &=&\frac{\kappa}{\sin^{2}\theta}\left[\kappa_{0}d^{l_2}_{m_2 0}(\theta) -\kappa_{1}d^{l_{2}+1}_{m_2 0}(\theta)  \right. \\
&& \ \left. - \ \kappa_{-1}d^{l_{2}-1}_{m_2 0}(\theta) +\kappa_{2}d^{l_{2}+2}_{m_2 0}(\theta)+\kappa_{-2}d^{l_{2}-2}_{m_2 0}(\theta) \right] \nonumber
\end{eqnarray}
and then as shown in \cite{SM-AS-TS}, the integrals for $m_2\neq0$ terms are obtained as 
\begin{widetext}
\begin{eqnarray}
I^{l_1 l_2}_{m_2,\pm2}\ &=&(-1)^{m_2}\sqrt{\frac{(2l_1 +1)}{4\pi}}\ \left\{
\begin{array}{ll}
\Big(\ \frac{\kappa \kappa_{0}}{|m_2|}+
\frac{\kappa \kappa_{2}}{|m_2|}\sqrt{\frac{(l_{2}+|m_2|)!(l_{2}+2-|m_2|)!}{(l_{2}-|m_2|)!(l_{2}+2+|m_2|)!}}\nonumber \\
+\frac{\kappa \kappa_{-2}}{|m_2|}\sqrt{\frac{(l_{2}-|m_2|)!(l_{2}-2+|m_2|)!}{(l_{2}+|m_2|)!(l_{1}-2-|m_2|)!}}\ \Big) & \mbox{if ($l_1 =l_2$)} \\ \\
\\ \\
\Big(\ \frac{\kappa \kappa_{0}}{|m_2|}\sqrt{\frac{(l_{2}+|m_2|)!(l_{1}-|m_2|)!}{(l_{2}-|m_2|)!(l_{1}+|m_2|)!}}+
\frac{\kappa \kappa_{2}}{|m_2|}\sqrt{\frac{(l_{2}+2+|m_2|)!(l_{1}-|m_2|)!}{(l_{2}+2-|m_2|)!(l_{1}+|m_2|)!}}
+\nonumber\\\frac{\kappa \kappa_{-2}}{|m_2|}\sqrt{\frac{(l_2-2+|m_2|)!(l_1-|m_2|)!}{(l_2-2-|m_2|)!(l_1+|m_2|)!}}\pm
\frac{\kappa \kappa_{1}}{|m_2|}\sqrt{\frac{(l_{2}+1+|m_2|)!(l_{1}-|m_2|)!}{(l_{2}+1-|m_2|)!(l_{1}+|m_2|)!}}
\nonumber\\\pm\frac{\kappa \kappa_{-1}}{|m_2|}\sqrt{\frac{(l_{2}-1+|m_2|)!(l_1-|m_2|)!}{(l_{2}-1-|m_2|)!(l_1+|m_2|)!}}\ \Big) & \mbox{if ($l_1 >l_2$)} \\ \\
\\ \\
\Big(\ \frac{\kappa \kappa_{0}}{|m_2|}\sqrt{\frac{(l_{1}+|m_2|)!(l_{2}-|m_2|)!}{(l_{1}-|m_2|)!(l_{2}+|m_2|)!}}+
\frac{\kappa \kappa_{2}}{|m_2|}\sqrt{\frac{(l_{2}+2-|m_2|)!(l_{1}+|m_2|)!}{(l_{2}+2+|m_2|)!(l_{1}-|m_2|)!}}\nonumber\\
+\frac{\kappa \kappa_{-2}}{|m_2|}\sqrt{\frac{(l_{2}-2-|m_2|)!(l_1+|m_2|)!}{(l_{2}-2+|m_2|)!(l_1-|m_2|)!}}\pm
\frac{\kappa \kappa_{1}}{|m_2|}\sqrt{\frac{(l_{2}+1-|m_2|)!(l_{1}+|m_2|)!}{(l_{2}+1+|m_2|)!(l_{1}-|m_2|)!}}
\nonumber\\ \pm\frac{\kappa \kappa_{-1}}{|m_2|}\sqrt{\frac{(l_{2}-1-|m_2|)!(l_1+|m_2|)!}{(l_{2}-1+|m_2|)!(l_1-|m_2|)!}}\ \Big) & \mbox{if ($l_1 <l_2$)}\,. \\ \\
\end{array}
\right.
\end{eqnarray}

In general, NC-beams would generate both even-parity and odd-parity
beam-BipoSH coefficients
\begin{eqnarray}\label{eq:oddparity-BipoSH}
B^{LM^{(NC)}}_{l_1 l_2}=B^{LM^{(+)}}_{l_1 l_2}+B^{LM^{(-)}}_{l_1 l_2}\,.
\end{eqnarray}

The even-parity beam-BipoSH,
\begin{eqnarray}
B^{LM^{(+)}}_{l_1 l_2}\ &=&\ \left\{ 
\begin{array}{ll}
 \delta_{M0}[b_{l_2 2}(\hat z)\exp({-i2\rho})+b^{*}_{l_2 2}(\hat z)\exp({i2\rho})](-C^{L0}_{l_1 0 l_2 0}\sqrt{\frac{4\pi l_2 (l_{2}-1)}
{(2l_{2}+1)(l_{2}+2)(l_{2}+1)}}+ \\2\pi\sqrt{\frac{(2l_{1}+1)}{4\pi}}\sum_{|m_2|>0}C^{L0}_{l_1 -m_2 l_2 m_2} \big[ \frac{\kappa \kappa_{0}}{|m_2|}+
\frac{\kappa \kappa_{2}}{|m_2|}\sqrt{\frac{(l_{2}+|m_2|)!(l_{2}+2-|m_2|)!}{(l_{2}-|m_2|)!(l_{2}+2+|m_2|)!}}\nonumber \\
+\frac{\kappa \kappa_{-2}}{|m_2|}\sqrt{\frac{(l_{2}-|m_2|)!(l_{2}-2+|m_2|)!}{(l_{2}+|m_2|)!(l_{1}-2-|m_2|)!}}\big]) & \mbox{if ($l_1 =l_2$ and $l_2\geq2$)} \\ \\

 \delta_{M0}[b_{l_2 2}(\hat z)\exp({-i2\rho})+b^{*}_{l_2 2}(\hat z)\exp({i2\rho})]2\pi\sqrt{\frac{(2l_{1}+1)}{4\pi}}(\sum_{|m_2|>0}C^{L0}_{l_1 -m_2 l_1 m_2}
\times\nonumber\\\big[ \frac{\kappa \kappa_{0}}{|m_2|}\sqrt{\frac{(l_{2}+|m_2|)!(l_{1}-|m_2|)!}{(l_{2}-|m_2|)!(l_{1}+|m_2|)!}}+
\frac{\kappa \kappa_{2}}{|m_2|}\sqrt{\frac{(l_{2}+2+|m_2|)!(l_{1}-|m_2|)!}{(l_{2}+2-|m_2|)!(l_{1}+|m_2|)!}}
+\nonumber\\\frac{\kappa \kappa_{-2}}{|m_2|}\sqrt{\frac{(l_2-2+|m_2|)!(l_1-|m_2|)!}{(l_2-2-|m_2|)!(l_1+|m_2|)!}}\big]) & \mbox{if ($l_1 >l_2$ and $l_2\geq2$)} \\ \\

\delta_{M0}[b_{l_2 2}(\hat z)\exp({-i2\rho})+b^{*}_{l_2 2}(\hat z)\exp({i2\rho})](8\pi C^{L0}_{l_1 0 l_2 0}\sqrt{\frac{(2l_{1}+1)(l_{2} -2)!}{4\pi(l_{2}+2)!}}+\nonumber\\2\pi\sqrt{\frac{(2l_{1}+1)}{4\pi}}\sum_{|m_2|>0}C^{L0}_{l_1 -m_2 l_1 m_2}\big[ \frac{\kappa \kappa_{0}}{|m_2|}\sqrt{\frac{(l_{1}+|m_2|)!(l_{2}-|m_2|)!}{(l_{1}-|m_2|)!(l_{2}+|m_2|)!}}+\nonumber\\
\frac{\kappa \kappa_{2}}{|m_2|}\sqrt{\frac{(l_{2}+2-|m_2|)!(l_{1}+|m_2|)!}{(l_{2}+2+|m_2|)!(l_{1}-|m_2|)!}}
+\frac{\kappa \kappa_{-2}}{|m_2|}\sqrt{\frac{(l_{2}-2-|m_2|)!(l_1+|m_2|)!}{(l_{2}-2+|m_2|)!(l_1-|m_2|)!}}\big]) & \mbox{if ($l_1 <l_2$ and $l_2\geq2$)}\,,
\end{array}
\right.
\end{eqnarray}
and odd parity-beam-BipoSH,
\begin{eqnarray}
B^{LM^{(-)}}_{l_1 l_2}\ &=&\ \left\{ 
\begin{array}{ll}
0 & \mbox{if ($l_1 = l_2$)} \\ \\
 \delta_{M0}[b_{l_2 2}(\hat z)\exp({-i2\rho})-b^{*}_{l_2 2}(\hat z)\exp({i2\rho})]2\pi\sqrt{\frac{(2l_{1}+1)}{4\pi}}(\sum_{|m_2|>0}C^{L0}_{l_1 -m_2 l_1 m_2}
\times\nonumber\\\big[
\frac{\kappa \kappa_{1}}{|m_2|}\sqrt{\frac{(l_{2}+1+|m_2|)!(l_{1}-|m_2|)!}{(l_{2}+1-|m_2|)!(l_{1}+|m_2|)!}}
+\frac{\kappa \kappa_{-1}}{|m_2|}\sqrt{\frac{(l_{2}-1+|m_2|)!(l_1-|m_2|)!}{(l_{2}-1-|m_2|)!(l_1+|m_2|)!}}\big]) & \mbox{if ($l_1 >l_2$ and $l_2\geq2$)} \\ \\

 \delta_{M0}[b_{l_2 2}(\hat z)\exp({-i2\rho})-b^{*}_{l_2 2}(\hat z)\exp({i2\rho})]2\pi\sqrt{\frac{(2l_{1}+1)}{4\pi}}(\sum_{|m_2|>0}C^{L0}_{l_1 -m_2 l_1 m_2}
\times\nonumber\\\big[
\frac{\kappa \kappa_{1}}{|m_2|}\sqrt{\frac{(l_{2}+1-|m_2|)!(l_{1}+|m_2|)!}{(l_{2}+1+|m_2|)!(l_{1}-|m_2|)!}}
+\frac{\kappa \kappa_{-1}}{|m_2|}\sqrt{\frac{(l_{2}-1-|m_2|)!(l_1+|m_2|)!}{(l_{2}-1+|m_2|)!(l_1-|m_2|)!}}\big]) & \mbox{if ($l_1 <l_2$ and $l_2\geq2$)} \\ \\
\end{array}
\right.
\end{eqnarray}
\end{widetext}

To avoid any confusion, we reiterate that the above results hold for
PT-scan approximation and a NC-beam function with discrete even-fold
azimuthal symmetry. Other residual symmetries in NC-beam can reduce
the set of non-zero beam BipoSH further.  In particular, if the
experimental beam has reflection symmetry, then odd parity beam BipoSH
will vanish and only even parity ones will be present. This implies
that odd parity beam BipoSH can be used as a measure of breakdown of
reflection symmetry in NC-beams.

\section{CMB ${\rm BipoSH}$ due to non-circular beams}\label{app:cmb-biposh}

The cosmological signal in the observed temperature fluctuations is
convolved with instrumental beam response function. So even if the
underlying cosmological temperature fluctuations are statistically
isotropic, non-circularity of the beam can give rise to detections in
BipoSH coefficients. In this appendix we provide a detailed derivation
of the BipoSH coefficient arising from NC beam presented in
Sec.~\ref{bipforbeam}. The measured temperature fluctuation map
$\widetilde{\Delta T}(\hat n_{1})$ is a convolution
\begin{eqnarray}\label{eq:convolution}
\widetilde{\Delta T}(\hat n_{1})=\int d{\Omega_{n_{2}}}B(\hat
n_{1},\hat n_{2})\Delta T(\hat n_{2}).
\end{eqnarray}
of the cosmological signal $\Delta T(\hat n_{2})$ with the beam
response function $B(\hat n_{1},\hat n_{2})$ that encodes the
sensitivity of the instrument around the pointing direction, $\hat
n_1$ and can be expanded in the spherical harmonic (SH) basis,
\begin{eqnarray}\label{eq:bctempfield}
\Delta \tilde T(\hat n)=\sum_{lm}\tilde a_{lm}Y_{lm}(\hat n).
\end{eqnarray}
Similarly, the cosmological signal decomposed in the SH basis, as
\begin{eqnarray}
\Delta T(\hat n) =\sum_{lm}a_{lm}Y_{lm}(\hat n)\,.
\end{eqnarray}
Beam response function can be expanded in the BipoSH basis,
\begin{eqnarray}
B(\hat n_{1},\hat n_{2})=\sum_{l_1 l_2 L M}B^{LM}_{l_1 l_2}\sum_{m_1 m_2}C^{LM}_{l_1 m_1 l_2 m_2}\times\nonumber\\ Y_{l_1 m_1}(\hat n_{1})Y_{l_1 m_1}(\hat n_{2}).
\end{eqnarray}
Using orthogonality of spherical harmonics,
\begin{eqnarray}
\int d\Omega_{\hat n} Y_{lm}(\hat n)Y_{l'm'}(\hat n)=(-1)^{m'}\delta_{ll'}\delta_{mm'}\,,
\end{eqnarray}
we obtain
\begin{equation}
\widetilde{\Delta T}(\hat n_{1})=\sum_{l_1 m_2}\sum_{l m L
M}(-1)^{m}a_{lm}B^{LM}_{l_1 l}C^{LM}_{l_1 m_1 l -m} Y_{l_1 m_1}(\hat
n_{1}).
\end{equation}
Using the above expansion together with Eq.~(\ref{eq:bctempfield}), we
obtain,
\begin{eqnarray}
\tilde a_{l_1 m_1}=\sum_{l m L M}(-1)^{m}a_{lm}B^{LM}_{l_1 l}C^{LM}_{l_1 m_1 l -m}.
\end{eqnarray}
and the harmonic space covariance as 
\begin{eqnarray}\label{eq:1}
&&\langle \tilde a_{l_1 m_1}\tilde a_{l_2 m_2}\rangle \ = \ \sum_{lmLM}\sum_{l'm'L'M'} (-1)^{m+m'}\langle a_{l m}a_{l'
m'}\rangle \nonumber\\
&&\quad \times \ B^{LM}_{l_1 l}B^{LM}_{l_2 l'}C^{LM}_{l_1 m_1 l
-m}C^{L'M'}_{l_2 m_2 l' -m'}\,.
\end{eqnarray}
Assuming the cosmological signal to be statistically isotropic,
\begin{eqnarray}
\langle a_{l m}a_{l' m'}\rangle=(-1)^{m}C_{l}\delta_{l l'}\delta_{m -m'}. 
\end{eqnarray}
and substituting in Eq.~(\ref{eq:1}), we obtain the SH-space covariance
of the observed map as
\begin{eqnarray}
&&\langle \tilde a_{l_1 m_1}\tilde a_{l_2 m_2}\rangle \ = \\
&& \quad \sum_{lmLM}\sum_{L'M'} (-1)^{m}C_{l} B^{LM}_{l_1 l}B^{L'M'}_{l_2 l}C^{LM}_{l_1 m_1 l -m}C^{L'M'}_{l_2 m_2 l m} \nonumber
\end{eqnarray}

As given in Eq.~(\ref{eq:gen-BipoSH}), CMB BipoSH coefficients are
related to the SH space covariance matrix in Eq.~(\ref{eq:1}), leading
to 
\begin{eqnarray}
\tilde A^{L_1 M_{1}}_{l_1 l_2} &=& \sum_{l L L' M M'} C_{l} B^{LM}_{l_1 l}B^{L'M'}_{l_2 l}\ \times\\
&&\quad\sum_{m m_1 m_2}(-1)^{m} C^{LM}_{l_1 m_1 l -m} C^{L' M'}_{l_2 m_2 l m} C^{L_1 M_1}_{l_1 m_1 l_2 m_2} \nonumber
\end{eqnarray}
The sum over product of three Clebsch-Gordan coefficients can be
written compactly in terms of a $6$-j symbol, as
\begin{equation}
\sum_{\alpha\beta\delta}(-1)^{a-\alpha}C^{c\gamma}_{a\alpha
b\beta}C^{e\epsilon}_{d\delta b\beta}C^{f\varphi}_{d\delta a
-\alpha}=K_1\prod_{cf}C^{e\epsilon}_{c\gamma f\varphi}
{\begin{Bmatrix} a & b & c \\ e & f & d
\end{Bmatrix}}\,,
\end{equation}
where $K_1 =(-1)^{b+c+d+f}$ and $\prod_{cf}=\sqrt{(2c+1)(2f+1)}$.  In
a PT-scan approximation, $M=0,M'=0, M_{1}=0$.  Hence, we obtain the
expression in Eq.~(\ref{biposhbeam}) for CMB BipoSH coefficient from
NC-beam with PT-scan approximation
 \begin{eqnarray}
&&\tilde A^{L_1 M_1}_{l_1 l_2} \ =\ \delta_{M_1 0} \sum_{lL L'} C_{l} \, B^{L 0}_{l_1 l}\, B^{L' 0}_{l_2 l} \, (-1)^{l_1+L'-L_{1}} \ \times\nonumber\\
&&\quad  \sqrt{(2L+1)(2L'+1)}C^{L_1 0}_{L 0 L' 0}
{\begin{Bmatrix}
l & l_1 & L \\
L_1 & L' & l_2  
\end{Bmatrix}}
\,.
\end{eqnarray}
The Clebsch-Gordan coefficient $C^{L_1 0}_{L 0 L' 0}$ is zero when the
sum $L+L'+L_{1}$ is odd valued, hence, enforces the condition that the
summation in the above expression is limited to $L+L'+L_{1}$ being
even-valued. When the beam function has an even fold azimuthal symmtery and reflection symmetry beam-BipoSH coefficients are restricted to even parity and follows 
$l_{1}+l_{2}=\textrm{even}$, then $L$ and $L'$ are restricted to even multipole values. Thereafter, due to the presence of $C^{L_1 0}_{L 0 L' 0}$, $L_1$ takes up even 
multipole values. 

\section{Effective averaging of beam due to multiple hits with varying orientations}
\label{beamavghits}

Here we have presented an analytic treatment to cover NC-beam effect
incorporating the multiple hits at any pixel $\hat{n}$ by the NC-beam
with fi{}xed shape but at varying orientations, $\rho_{j}(\hat{n})$.
This information can be obtained from the instrument design description
and the scan-strategy of the experiment over the duration of the data
acquisition. 

The observed temperature anisotropy for a single hit by the beam at
a direction $\hat{n}\equiv(\theta,\phi)$
with orientation, $\rho_{j}(\hat{n})$ is given by

\begin{eqnarray}
T(\gamma) & = & \int d\Omega_{\hat{n}'}B\left(\hat{n},\hat{n}';\rho(\hat{n})\right)T(\hat{n}')\nonumber \\
 & = & \int d\Omega_{\hat{n}'}\left(\sum b_{lm}\left(\hat{n},\rho(\hat{n})\right)Y_{lm}(\hat{n}')\right)T(\hat{n}')\nonumber \\
 & = & \sum b_{lm}\left(\hat{n},\rho(\hat{n})\right)\int T(\hat{n}')Y_{lm}(\hat{n}')d\Omega_{\hat{n}'}\nonumber \\
 & = & \sum a_{lm}b_{lm}\left(\hat{n},\rho(\hat{n})\right)
\end{eqnarray}

If the $i^{th}$ pixel gets scanned $n_{i}$ number of times an approximate
observed temperature $T_{s}(\gamma)$ of that pixel is given by 

\begin{eqnarray}
T_{s}(\gamma) & = & \frac{1}{n_{i}}\sum_{j=1}^{n_{i}}\sum_{lm}a_{lm}b_{lm}\left(\hat{n},\rho_{j}(\hat{n})\right) \\
 & = & \frac{1}{n_{i}}\sum_{j=1}^{n_{i}}\sum_{lm}a_{lm}\sum_{m'}b_{lm'}(\hat{z})d_{mm'}^{l}(\theta)e^{-im\phi}e^{-im'\rho_{j}} \nonumber
\end{eqnarray}

Of course, if a pixel get hit $n_{i}$times from different orientations
then after following the map making procedure for the differential
assembly, its temperature may not be average of all the hits. But
the average can provide a very good estimate of the scanned temperature
when there is no noise involved and also simplifies the calculation.
Here $\hat{z}$ is a fixed direction in the sky.

Not if we follow the parallel transport with $\rho=0$, then the temperature
of any pixel will be given by

\begin{eqnarray}
T_{pt}(\gamma) & = & \sum_{lm}a_{lm}b_{lm}^{effective}\left(\hat{n},0\right)\\
 & = & \sum_{lm}a_{lm}\sum_{m'}b_{lm'}^{effective}(\hat{z})d_{mm'}^{l}(\theta)e^{-im\phi}
\end{eqnarray}

Now we want to choose the effective beam in such a way that the error
in the temperature calculation be minimum. In other words we want
to minimize the following factor. 

\begin{widetext}
\begin{eqnarray}
\chi^{2} & = & \int d\Omega\left(\frac{1}{n_{i}}\sum_{j=1}^{n_{i}}\sum_{lm}a_{lm}\sum_{m'}b_{lm'}(\hat{z})d_{mm'}^{l}(\theta)e^{-im\phi}e^{-im'\rho_{j}}-\sum_{lm}a_{lm}\sum_{m'}b_{lm'}^{effective}(\hat{z})d_{mm'}^{l}(\theta)e^{-im\phi}\right)^{2}\\
 & = & \sum_{lm}a_{lm}^{2}\int d\Omega\left(\frac{1}{n_{i}}\sum_{j=1}^{n_{i}}\sum_{m'}b_{lm'}(\hat{z})d_{mm'}^{l}(\theta)e^{-im'\rho_{j}}-\sum_{m'}b_{lm'}^{\text{eff}}(\hat{z})d_{mm'}^{l}(\theta)\right)^{2}e^{-2im\phi}
\end{eqnarray}

As we want to minimize this error by choosing a effective $b_{lm}^{\text{eff}}$,
we have to take the derivative of $\chi^{2}$ with respect to effective
$b_{lm}^{\text{eff}}$ and make it $0$. This gives 

\begin{equation}
\frac{\partial\chi^{2}}{\partial b_{lm'}^{\text{eff}}}=\sum_{lm}a_{lm}^{2}e^{-2im\phi}\int d\Omega\left(\frac{1}{n_{i}}\sum_{i}\sum_{m'}b_{lm'}(\hat{z})d_{mm'}^{l}(\theta)e^{-im'\rho_{i}}-\sum_{m''}b_{lm''}^{\text{eff}}(\hat{z})d_{mm''}^{l}(\theta)\right)d_{mm''}^{l}(\theta)
\end{equation}

and hence 

\begin{equation}
2\sum_{lm}a_{lm}^{2}e^{-2im\phi}\int d\Omega\left(\frac{1}{n_{i}}\sum_{i}\sum_{m'}b_{lm'}(\hat{z})d_{mm'}^{l}(\theta)e^{-im'\rho_{i}}-\sum_{m''}b_{lm''}^{\text{eff}}(\hat{z})d_{mm''}^{l}(\theta)\right)d_{mm''}^{l}(\theta)=0
\end{equation}

As $a_{lm}$ are random quantities, to make the sum $0$, each and
every coefficients which are with $a_{lm}^{2}$ should be $0$. This
means for all the $l$ and $m$'s we must have 

\begin{equation}
\int d\Omega\left(\frac{1}{n_{i}}\sum_{i}\sum_{m'}b_{lm'}(\hat{z})d_{mm'}^{l}(\theta)e^{-im'\rho_{i}}-\sum_{m''}b_{lm''}^{\text{eff}}(\hat{z})d_{mm''}^{l}(\theta)\right)d_{mm''}^{l}(\theta)=0
\end{equation}

The integration over $d\Omega$ can be replaced by the summation over
all the pixels. i.e.

\begin{equation}
\sum_{N_{p}}\left(\frac{1}{n_{i}}\sum_{i}\sum_{m'}b_{lm'}(\hat{z})d_{mm'}^{l}(\theta)e^{-im'\rho_{i}}-\sum_{m'}b_{lm''}^{\text{eff}}(\hat{z})d_{mm''}^{l}(\theta)\right)d_{mm''}^{l}(\theta)=0
\end{equation}

\begin{equation}
\Rightarrow\sum_{N_{p}}\frac{1}{n_{i}}\sum_{i}\sum_{m'}b_{lm'}(\hat{z})d_{mm'}^{l}(\theta)d_{mm''}^{l}(\theta)e^{-im'\rho_{i}}=\sum_{N_{p}}\sum_{m'}b_{lm'}^{\text{eff}}(\hat{z})d_{mm''}^{l}(\theta)d_{mm'}^{l}(\theta)
\end{equation}
\end{widetext}

here the summation over $N_{p}$ runs over all the pixels.
Few straightforward algebraic manipulation will give us 
\begin{equation}
b_{l2}^{\text{eff}}= \frac{2\, l + 1}{N_p} \sum_{N_{p}}\frac{1}{n_{i}}\sum_{i}\sum_{m'}b_{lm'}(\hat{z})d_{2m'}^{l}(\theta)d_{2m'}^{l}(\theta)e^{-im'\rho_{i}} \, .
\end{equation}

Now if in the real beam if we consider that the $m'=2$ part is only
the important part then we can get 

\begin{equation}\label{eq:beamavghits}
b_{l2}^{\text{eff}} \ = \ \frac{2\, l + 1}{N_p} b_{l2}(\hat{z})\sum_{N_{p}}\frac{1}{n_{i}}\sum_{i}d_{22}^{l}(\theta)d_{22'}^{l}(\theta)e^{-im'\rho_{i}} \, .
\end{equation}

If we do the calculation using the WMAP scan strategy, it can be seen
that this factor with $b_{l2}(\hat{z})$ is almost constant over $l$
with a value $\sim0.45$ (refer to Fig.(\ref{fig:beamavghits})). From the plots we can also see that the
constant multiplication which is coming is $\sim0.45$ (refer to Fig.(\ref{fig:eff-BipoSH})), which is very
close to the factor calculated here. So we can say that doing only
a time efficient parallel transport scan also it s possible to calculated
an approximate effect the beam induced BipoSH.

\end{document}